\documentclass[floatfix, showkeys, reprint, superscriptaddress, amsmath, amssymb, aps]{revtex4-1}

\usepackage{graphicx}
\usepackage{dcolumn}
\usepackage{bm}
\usepackage{subfigure}
\usepackage{caption}
\usepackage{subcaption}
\usepackage[caption=false]{subfig}
\usepackage{float}
\usepackage{booktabs}
\usepackage{multirow}
\usepackage[normalem]{ulem}
\usepackage{soul}
\usepackage{xcolor}
\usepackage{ragged2e}

\begin{document}

\preprint{APS/123-QED}

\title{Exploring Quantum Machine Learning for Weather Forecasting}

\author{Maria Heloísa F. da Silva}
\email{maria.fraga@fbter.org.br}
\affiliation{QuIIN—Quantum Industrial Innovation, EMBRAPII CIMATEC Competence Center in Quantum Technologies, SENAI CIMATEC, Av. Orlando Gomes, 1845, Salvador, Bahia, 41850-010, Brazil}
\affiliation{Quantum Information Group, Federal University of Western Bahia - Campus Reitor Edgard Santos, Rua Bertioga, 892, Morada Nobre I, Barreiras, Bahia, 47810-059, Brazil}

\author{Gleydson F. de Jesus}
\email{gleydson.jesus@fieb.org.br}
\affiliation{QuIIN—Quantum Industrial Innovation, EMBRAPII CIMATEC Competence Center in Quantum Technologies, SENAI CIMATEC, Av. Orlando Gomes, 1845, Salvador, Bahia, 41850-010, Brazil}

\author{Christiano M. S. Nascimento}
\email{christiano.moreira@fbter.org.br}
\affiliation{QuIIN—Quantum Industrial Innovation, EMBRAPII CIMATEC Competence Center in Quantum Technologies, SENAI CIMATEC, Av. Orlando Gomes, 1845, Salvador, Bahia, 41850-010, Brazil}

\author{Valéria L. da Silva}
\email{valeria.dasilva@fieb.org.br}
\affiliation{QuIIN—Quantum Industrial Innovation, EMBRAPII CIMATEC Competence Center in Quantum Technologies, SENAI CIMATEC, Av. Orlando Gomes, 1845, Salvador, Bahia, 41850-010, Brazil}

\author{Clebson S. Cruz}
\email{clebson.cruz@ufob.edu.br}
\affiliation{Quantum Information Group, Federal University of Western Bahia - Campus Reitor Edgard Santos, Rua Bertioga, 892, Morada Nobre I, Barreiras, Bahia, 47810-059, Brazil}

\date{\today}

\begin{abstract}
Weather forecasting plays a crucial role in supporting strategic decisions across various sectors, including agriculture, renewable energy production, and disaster management. However, the inherently dynamic and chaotic behavior of the atmosphere presents significant challenges to conventional predictive models. On the other hand, introducing quantum computing simulation techniques to the forecasting problems constitutes a promising alternative to overcome these challenges. In this context, this work explores the emerging intersection between quantum machine learning (QML) and climate forecasting. We present the implementation of a Quantum Neural Network (QNN) trained on real meteorological data from NASA’s Prediction of Worldwide Energy Resources (POWER) database. The results show that QNN has the potential to outperform a classical Recurrent Neural Network (RNN) in terms of accuracy and adaptability to abrupt data shifts, particularly in wind speed prediction. Despite observed nonlinearities and architectural sensitivities, the QNN demonstrated robustness in handling temporal variability and faster convergence in temperature prediction. These findings highlight the potential of quantum models in short- and medium-term climate prediction, while also revealing key challenges and future directions for optimization and broader applicability.
\end{abstract}

\maketitle

\section{Introduction}\label{sec1}

The atmosphere is a dynamic and chaotic system, in which slight variations in initial conditions can lead to significant changes in future outcomes \cite{shen2022dual}. In this context, weather forecasting is the process of estimating future climate conditions for a given location based on the analysis of historical climate data. Historically, climate forecasting has contributed to the advancement of humanity in various domains, playing a crucial role in guiding actions in areas such as agriculture \cite{pujahari2022intelligent}, aeronautics \cite{schultz2021predictive, wang2021prediction}, tourism \cite{li2018big}, urban planning \cite{enriquez2016solar}, disaster management involving extreme weather events \cite{braman2013climate}, and the monitoring of climate phenomena \cite{van2021defining}. Additionally, climate studies are crucial for optimizing the performance of renewable energy generation systems: wind analysis can enhance the efficiency of wind turbines, while solar radiation studies can maximize solar energy production \cite{bochenek2022machine}.

Traditionally, this task has been guided by a wide range of climate models that simulate the behavior of the Earth's atmosphere, including physical, statistical, artificial intelligence (AI)-based, and hybrid models \cite{meenal2022weather}. However, despite their historical relevance, traditional models face paramount limitations. Among the primary difficulties are the need for approaches capable of processing large volumes of historical data with numerous interdependent climatic variables, the challenge of predicting rare events, and the limitations of traditional models in capturing long-term patterns effectively \cite{spiridonov2021weather, brotzge2023challenges}. 

In light of these challenges, technological advancements have paved the way for exploring novel approaches to enhance forecasting capabilities. In particular, the combination of quantum computing and machine learning has given rise to quantum machine learning (QML) \cite{zeguendry2023quantum}, an emerging field of artificial intelligence and a promising solution to this issue. The advantages of quantum models stem from quantum properties such as superposition and entanglement \cite{gill2022quantum, lins2024quantum}, which enable greater processing capacity and the identification of complex patterns that might otherwise remain undetected by classical computing models \cite{cerezo2022challenges}. Currently, most analyses in this field are stochastic, and the optimization of quantum neural architectures for improved results is still under investigation \cite{jaderberg2024potential}. 

Quantum machine learning has been mainly applied to classification tasks and has seen limited use in regression problems \cite{hong2023robust}. In this context, quantum machine learning for weather forecasting has received limited attention, with existing studies focusing on predicting wind speed \cite{blazakis2023one, hong2023day, hong2024hybrid}, solar irradiance \cite{sushmit2023forecasting, oliveira2024application, hong2024solar}, wind and photovoltaic power generation \cite{kou2020generation, guilian2023short}, geomagnetic storms \cite{alomari2023hybrid}, evapotranspiration \cite{mostafa2023modeling}, and precipitation \cite{deepamarine}.

This work proposes the implementation and performance analysis of a Quantum Neural Network (QNN) for weather forecasting, applied to real climate data. The main contribution of this paper lies in extending QML beyond its traditional application in classification to regression tasks, specifically for predicting temperature on a medium-term basis and wind speed on a short-term basis. The computational cost of a quantum circuit is typically assessed by the number of trainable parameters and the number of two-qubit operations \cite{mahmud2024quantum}. Focusing on this aspect, we varied the circuit depth exclusively by adjusting the variational layers to identify the configuration that maximizes predictive performance. Furthermore, the study includes an error analysis and a comparative evaluation of performance metrics against those obtained with a classical neural network trained on the same dataset.

\section{Material and Methods}\label{sec2}

This study followed a three-step methodology: selection of the meteorological dataset (\ref{basemet}), data preprocessing (\ref{tratamento}), and definition of machine learning models (\ref{modelos}). The choice of dataset is critical, as the quality and completeness of the data directly affect the performance of machine learning models \cite{gong2023survey, budach2022effects}.

\subsection{Meteorological Dataset Selection}
\label{basemet}

The analyzed region was the Brazilian city of Barreiras, located in the state of Bahia. This municipality plays a strategic role in national agribusiness, ranking 25th in agricultural production \cite{agricultura2023}, and hosts the Federal University of Western Bahia, which serves as a regional center for research and technological development. Geographically, it is positioned at approximately 12.15°S latitude and 44.99°W longitude \cite{ibge2020}, a location characterized by a tropical climate with marked wet and dry seasons, making it a relevant setting for weather forecasting studies.

Initially, we evaluate three candidate meteorological datasets: NASA Prediction of Worldwide Energy Resources (POWER) \cite{NASAPOWER}, the National Institute of Meteorology (INMET) \cite{INMET}, and HidroWeb - National Water Agency (ANA) \cite{Hidroweb}. As summarized in Table \ref{tab:datasets}, NASA POWER provided the most comprehensive and consistent dataset, with minimal missing data and extensive parameter coverage, which supports its selection for this study.

\begin{table}[h!]
\centering
\caption{ \justifying~  Meteorological parameters available in the evaluated datasets NASA Prediction of Worldwide Energy Resources (POWER) \cite{NASAPOWER}, the National Institute of Meteorology (INMET) \cite{INMET}, and HidroWeb - National Water Agency (ANA) \cite{Hidroweb}.}
\begin{tabular}{lccc}
\toprule
\textbf{Parameter} & \textbf{NASA} & \textbf{INMET} & \textbf{HidroWeb} \\
\midrule
Instantaneous temperature & Yes & Yes & Yes \\
Maximum temperature & Yes & Yes & No \\
Minimum temperature & Yes & Yes & No \\
Relative humidity & Yes & Yes & No \\
Dew point & Yes & Yes & No \\
Atmospheric pressure & Yes & Yes & Yes \\
Wind speed & Yes & Yes & Yes \\
Maximum wind speed & Yes & No & No \\
Minimum wind speed & Yes & No & No \\
Wind direction & Yes & Yes & Yes \\
Solar radiation & Yes & Yes & No \\
Precipitation & Yes & Yes & Yes \\
Solar fluxes & Yes & No & No \\
Soil properties & Yes & No & No \\
\bottomrule
\end{tabular}
\label{tab:datasets}
\end{table}

\subsection{Data Preprocessing}
\label{tratamento}

\begin{figure*}[t]
\centering
\includegraphics[scale=0.44]{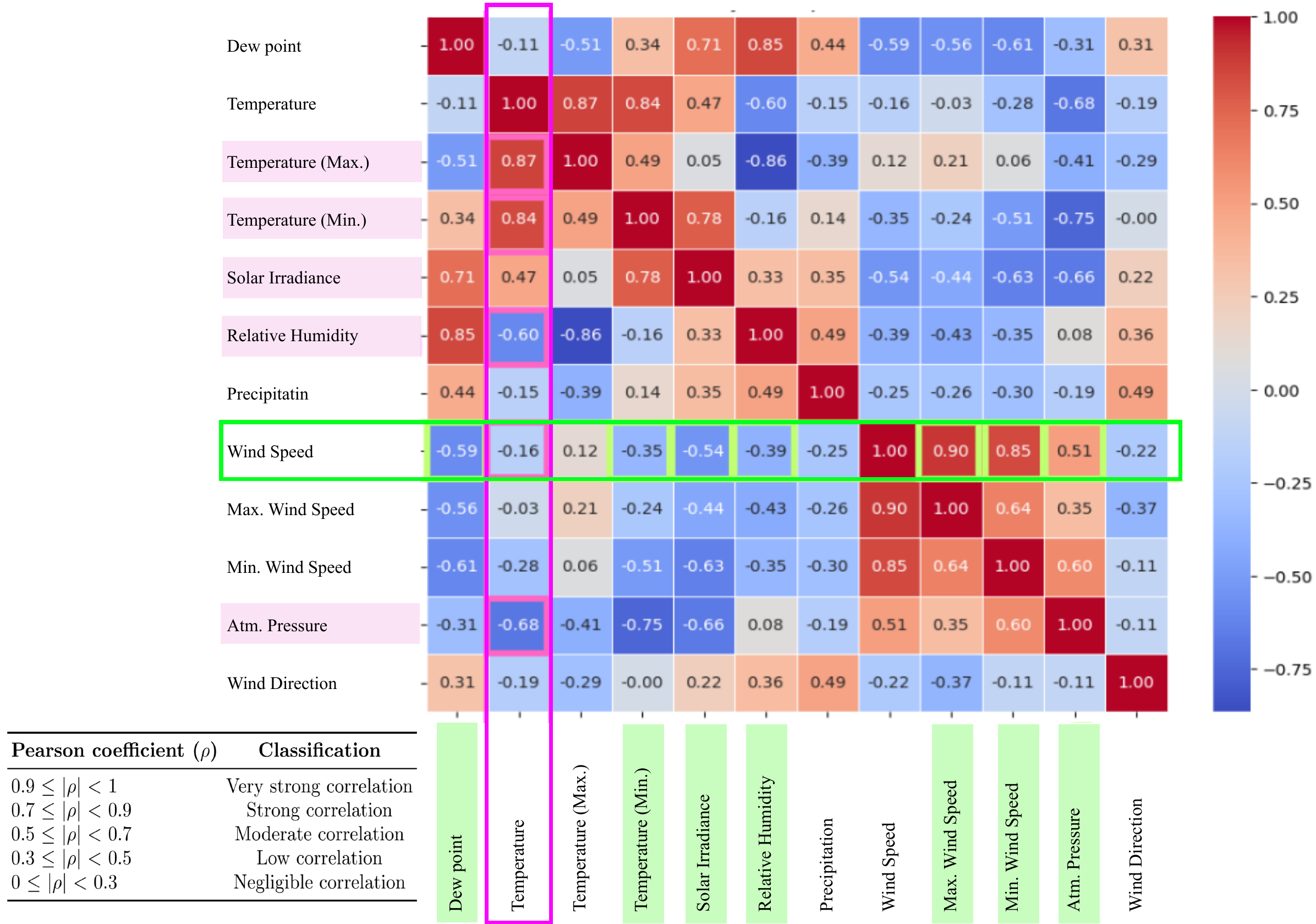}
\caption{ \justifying~Pearson correlation matrix for the selected meteorological parameters. The interpretation of the Pearson coefficient ($\rho$) is detailed in the inset table (bottom left). Strong positive correlations are indicated in intense red, while strong negative correlations are indicated in intense blue. The highlighted parameters represent the significant variables ($|\rho| > 0.3$) for the  target variables - magenta for the temperature and green for wind speed.}
\label{fig:correlation_matrix}
\end{figure*}

Preprocessing included feature selection and the creation of time-lagged variables to capture temporal dependencies. Feature selection was guided by the Pearson correlation matrix displayed in Figure \ref{fig:correlation_matrix}. The interpretation of the Pearson coefficient ($\rho$) is detailed in the inset table in Figure \ref{fig:correlation_matrix} (bottom left). To reduce computational costs associated with training and implementing QNN, features with negligible correlation ($|\rho| < 0.3$) with the targets were excluded from the dataset, retaining only variables with significant correlations to the target variables (temperature and wind speed).

\begin{figure}[h!]
\centering
\includegraphics[scale=0.45]{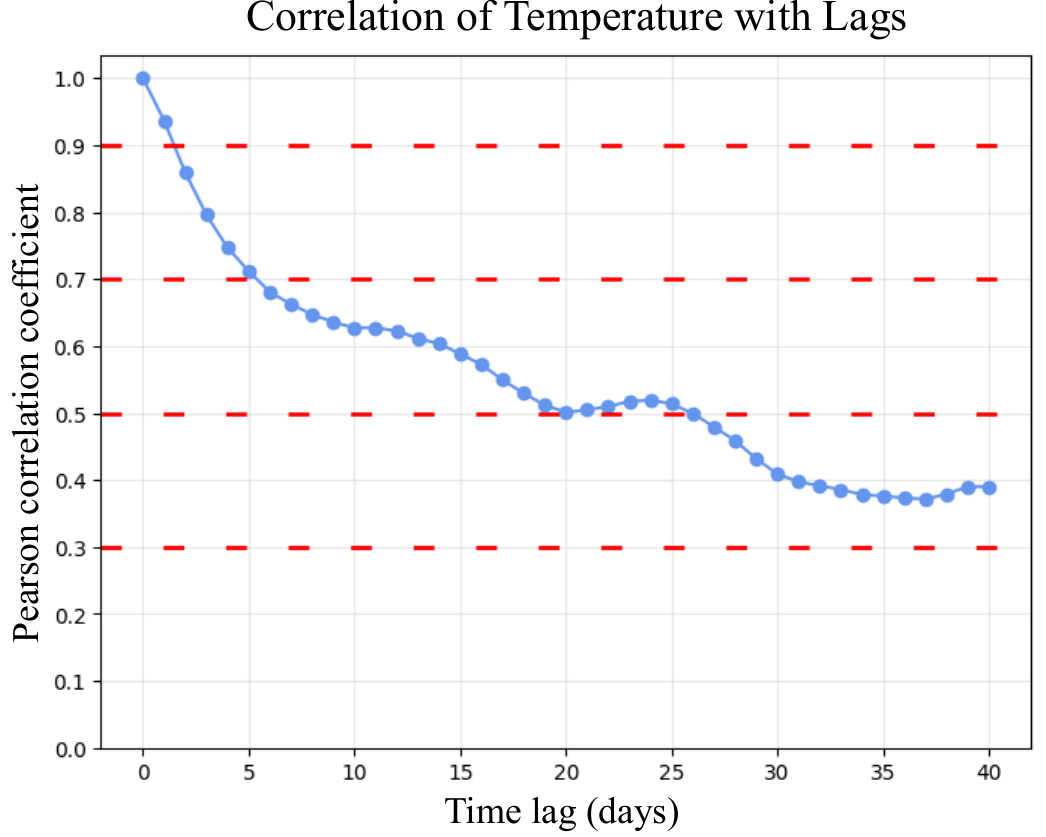} 
\caption{ \justifying~Pearson correlation between temperature and its time-lagged values over a 40-day period in Barreiras, Brazil.}
\label{fig:timelag_temp}
\end{figure}

\begin{figure}[h!]
\centering
\includegraphics[scale=0.45]{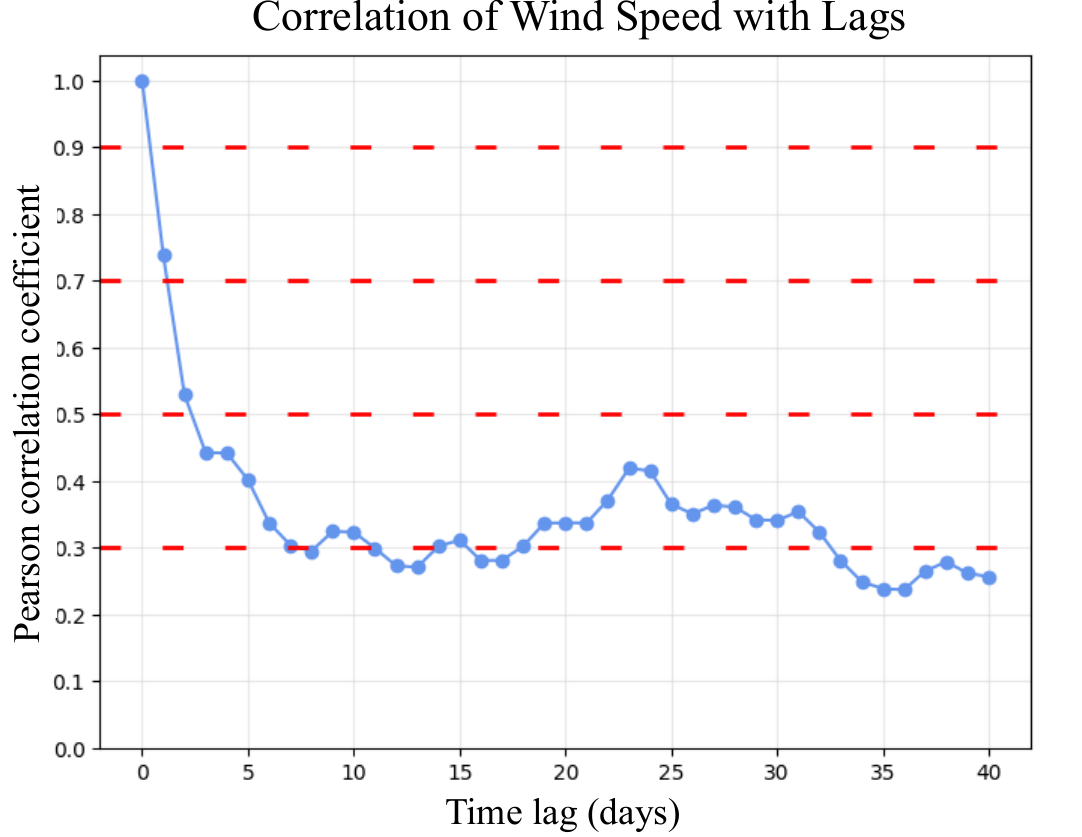} 
\caption{ \justifying~~Pearson correlation between wind speed and its time-lagged values over a 40-day period in Barreiras, Brazil.}
\label{fig:timelag_wind}
\end{figure}

Time-lagged variables were incorporated to account for temporal dependencies. Optimal lag periods were determined by computing Pearson correlations between each target variable and its lagged values. Figures \ref{fig:timelag_temp} and \ref{fig:timelag_wind} illustrate these correlations for temperature and wind speed, respectively.

Based on this analysis, lags of 28 days for temperature and 6 days for wind speed were appended to the dataset, yielding the final datasets used for model training. The statistical properties of the forecast target variables are summarized in Table \ref{tab:target_stats}.

\begin{table}[h!]
\centering
\caption{ \justifying~  Statistical properties of the forecast target variables.}
\begin{tabular}{lcccc}
\toprule
\textbf{Parameter} & \textbf{Mean} & \textbf{Std. Dev.} & \textbf{Min} & \textbf{Max} \\
\midrule
Temperature (°C) & 26.61 & 2.61 & 20.90 & 33.14 \\
Wind speed (m/s) & 2.03 & 0.67 & 0.64 & 4.56 \\
\bottomrule
\end{tabular}
\label{tab:target_stats}
\end{table}

After feature reduction, the data were standardized as

\begin{equation}
\hat{x} = \frac{x - \mu}{\sigma}\text{,}
\end{equation}

\noindent where $x$ is the original value, $\mu$ the mean, $\sigma$ the standard deviation, and $\hat{x}$ the standardized value \cite{ogur2023effect}. This procedure assumes an approximately normal data distribution and ensures all features are on a comparable scale, reducing bias and facilitating model interpretation \cite{de2025exploring}.

\subsection{Machine Learning Models}
\label{modelos}

\subsubsection{Quantum Neural Network (QNN)}

\begin{figure*}[!t]
\centering
\includegraphics[width=0.85\linewidth]{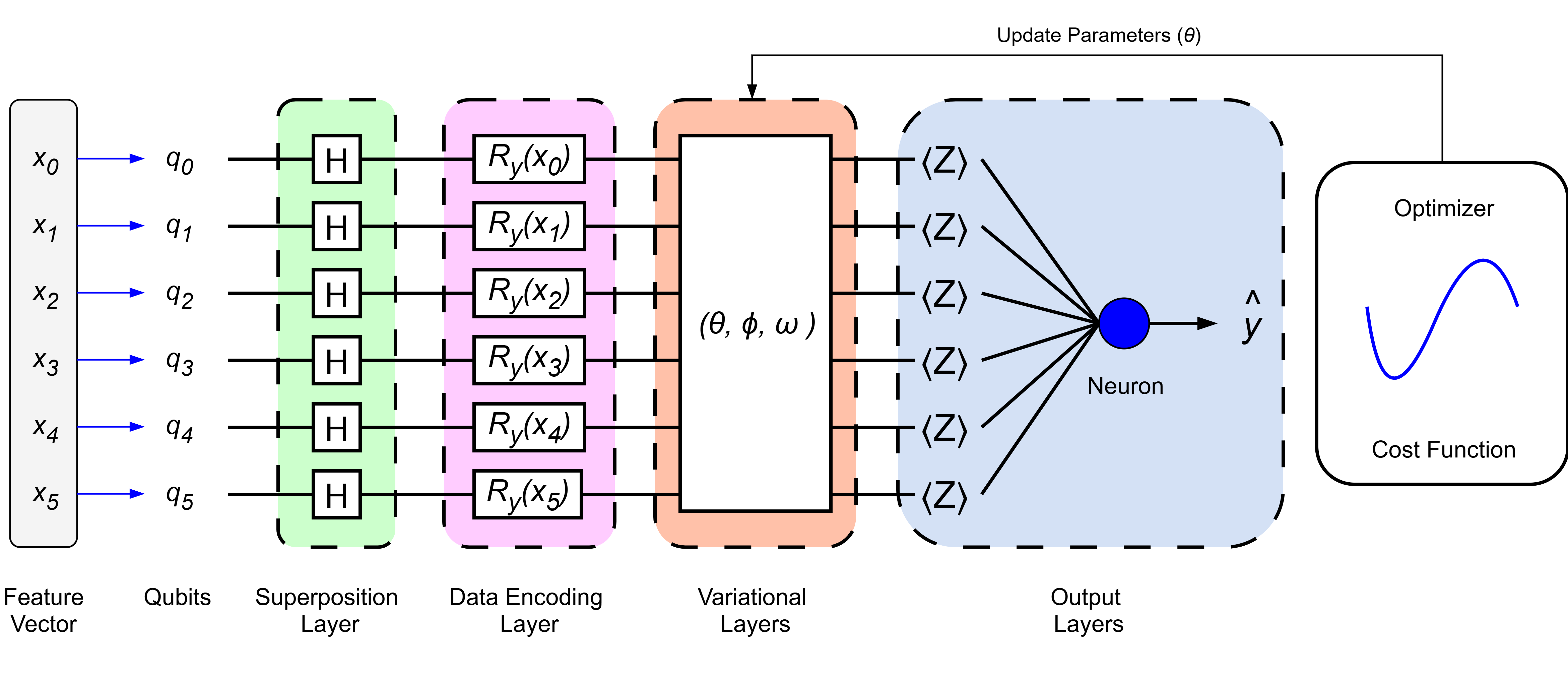} 
\caption{ \justifying~Operating principle of the variational quantum algorithm (VQA). Classical inputs—here, temperature and wind speed—denoted $x_i$ are (a) prepared in superposition and (b) encoded into a quantum state $\lvert \Psi_{\text{input}}(x_i)\rangle$ via unitary rotation gates $R_y(x_i)$. (c) Parameterized gates $R(\boldsymbol{\theta})$ assign trainable weights $\boldsymbol{\theta}$. (d) Projective measurement collapses the qubits to classical outcomes, which are then (e) fed to a classical optimizer that minimizes a cost function (prediction error) during training.}
\label{fig:vqa-principle}
\end{figure*}

\begin{figure}[!t]
\centering
\includegraphics[width=\linewidth]{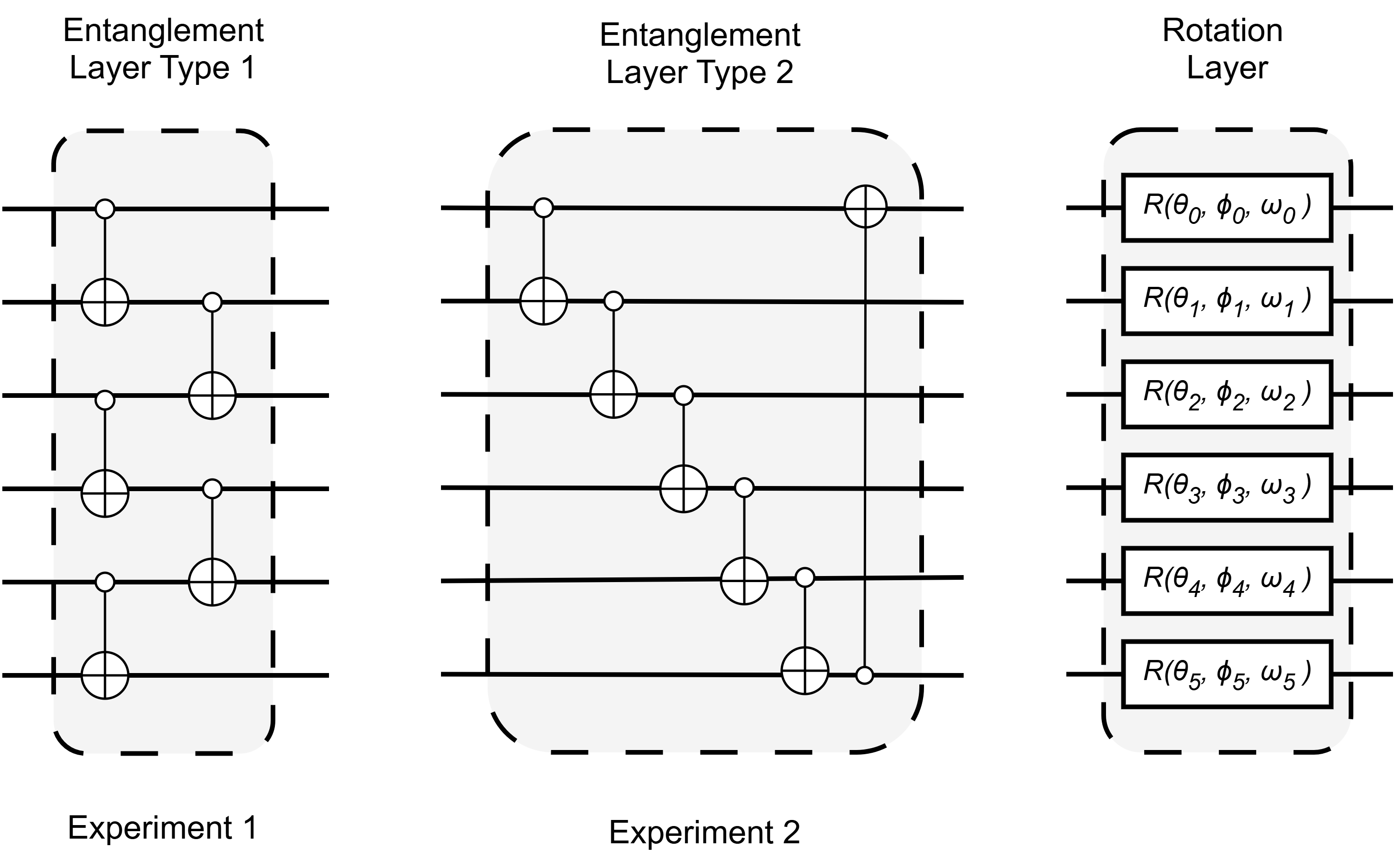} 
\caption{ \justifying~Entanglement strategies applied to the presented INN: (left) \textit{EntanglingLayer} (Experiment~1) and (center) \textit{StronglyEntanglingLayer} (Experiment~2). To assess the effect of network depth, each strategy is instantiated with $d\in\{1,3,5\}$ stacked variational layers, followed (right) by a layer of parameterized single-qubit rotations $R(\theta_i,\varphi_i,\omega_i)$. The combination of two entanglement types and three depths yields six QNN configurations in total.}
\label{fig:qnn-layers}
\end{figure}

The quantum predictive model employed in this study is a Quantum Neural Network (QNN), designed following the approach proposed by Oğur~\cite{ogur2023effect} and further explored by Jesus~\cite{de2025exploring} in demand forecasting for the financial sector. The operating principle of the underlying variational quantum algorithm—where classical features $x_i$ are encoded via $R_y(x_i)$ into $\lvert\Psi_{\text{input}}(x_i)\rangle$, processed by parameterized gates $R(\boldsymbol{\theta})$, measured, and then updated by a classical optimizer—is illustrated in Fig.~\ref{fig:vqa-principle}. To assess the influence of network depth on learning performance, architectures with 1, 3, and 5 variational layers were considered. Furthermore, two distinct entanglement strategies were evaluated, namely the \textit{EntanglingLayer} (Experiment~1) and the \textit{StronglyEntanglingLayer} (Experiment~2)~\cite{hong2023robust}, leading to a total of six QNN configurations, as depicted in Fig.~\ref{fig:qnn-layers}.

\subsubsection{Recurrent Neural Network (RNN)}

The classical predictive model adopted in this study was a Recurrent Neural Network (RNN), implemented following the approach described by \cite{ekman2021learning}. While this architecture has demonstrated effectiveness in time-series forecasting tasks, its ability to capture abrupt temporal variations is comparatively limited \cite{li2023quantum}.






\section{Results and Discussion}\label{sec3}

The temperature forecasting results are presented in Section \ref{subsec1} and are displayed according to the outputs of the QNN model (Section \ref{subsec1_q}) and the RNN model (Section \ref{subsec1_c}). Similarly, the results for wind speed forecasting are shown in Section \ref{subsec1} and organized according to the outputs of the Quantum Neural Network (QNN) model (Section \ref{subsec2_q}) and the Recurrent Neural Network (RNN) model (Section \ref{subsec2_c}).

Performance metrics are reported as statistical distributions, accompanied by the corresponding training and validation loss profiles for the previously described QNN and RNN architectures. Additionally, comparative tables and figures reporting the Mean Absolute Error (MAE) are included to provide a comprehensive evaluation of predictive performance. The experimental configuration parameters of the adapted QNN models for the target variable are also presented to ensure reproducibility.

\subsection{Temperature prediction}\label{subsec1}

\begin{figure}[H]
    \centering
    \includegraphics[width=0.9\linewidth]{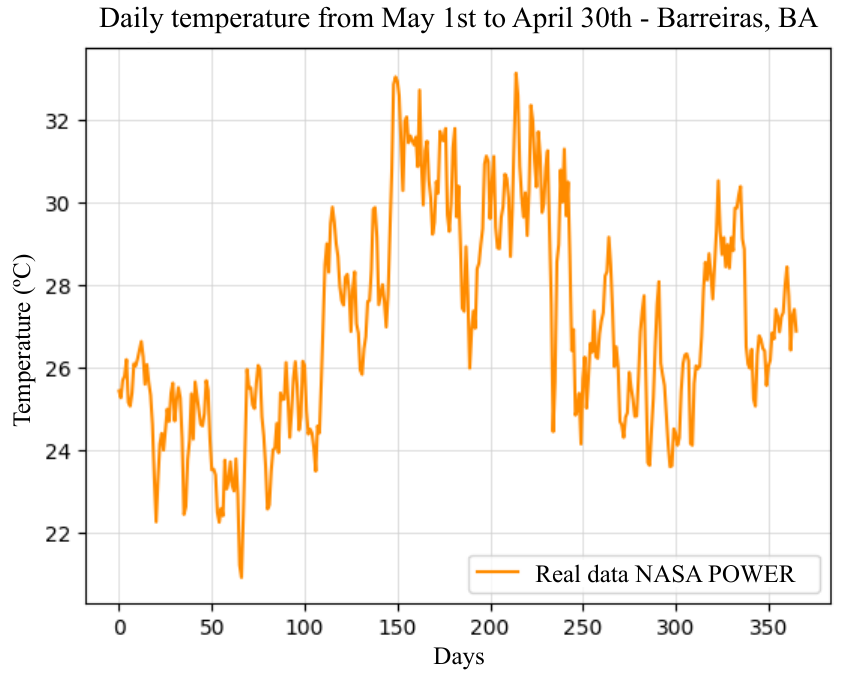}
    \caption{ \justifying~Actual daily temperature data obtained from the NASA POWER dataset.}
    \label{fig:dadosreaistemp}
\end{figure}

The temperature forecasting experiments were initially performed using a train-test split based on the defined time lag of 28 days. The prediction horizon was set to 14 days, representing the test set. Considering the dataset spans one year (from May 1st, 2023, to April 30th, 2024), the data were partitioned into training and testing subsets corresponding to 96.2\% and 3.8\%, respectively. Figure \ref{fig:dadosreaistemp} presents the daily temperature data for the specified period and location, while Figure \ref{fig:conjuntosvento} illustrates the previously calculated test and training proportions.


\subsubsection{Quantum Model} \label{subsec1_q}

\begin{figure*}[!t]
    \centering

    \begin{minipage}{0.48\textwidth}
        \centering
        \includegraphics[width=0.9\linewidth]{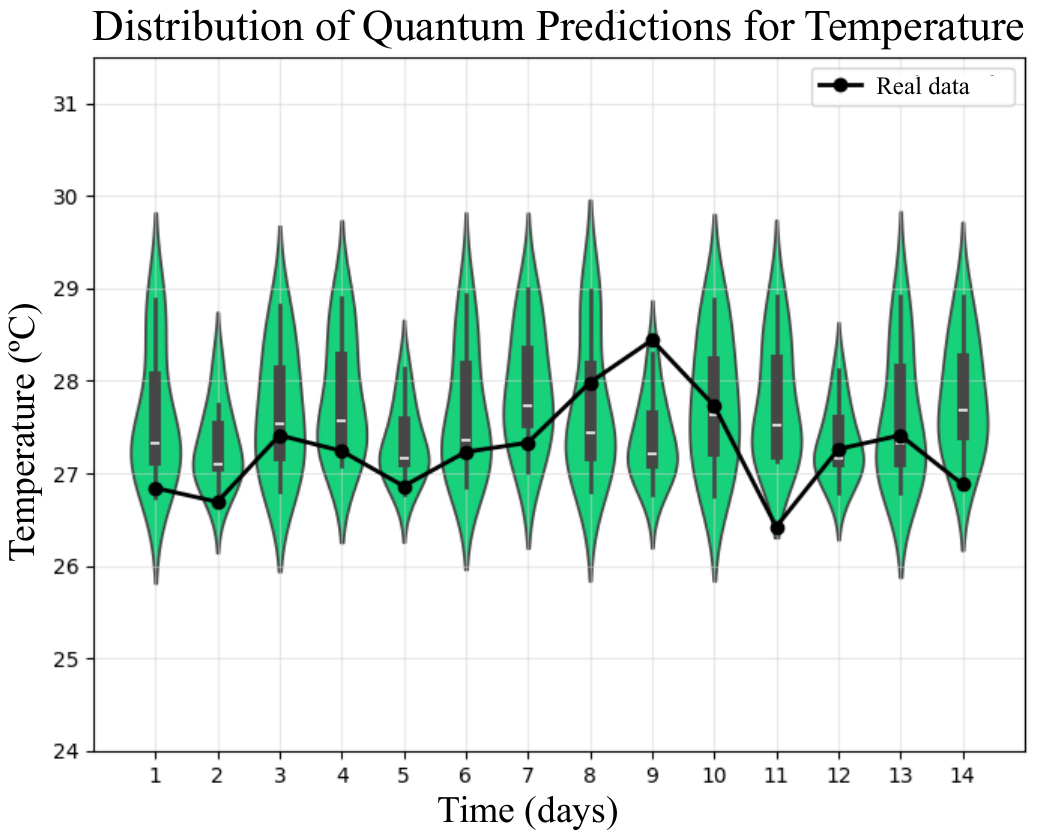}
        \caption*{Exp 1 – 1 layer}
    \end{minipage}
    \hfill
    \begin{minipage}{0.48\textwidth}
        \centering
        \includegraphics[width=0.9\linewidth]{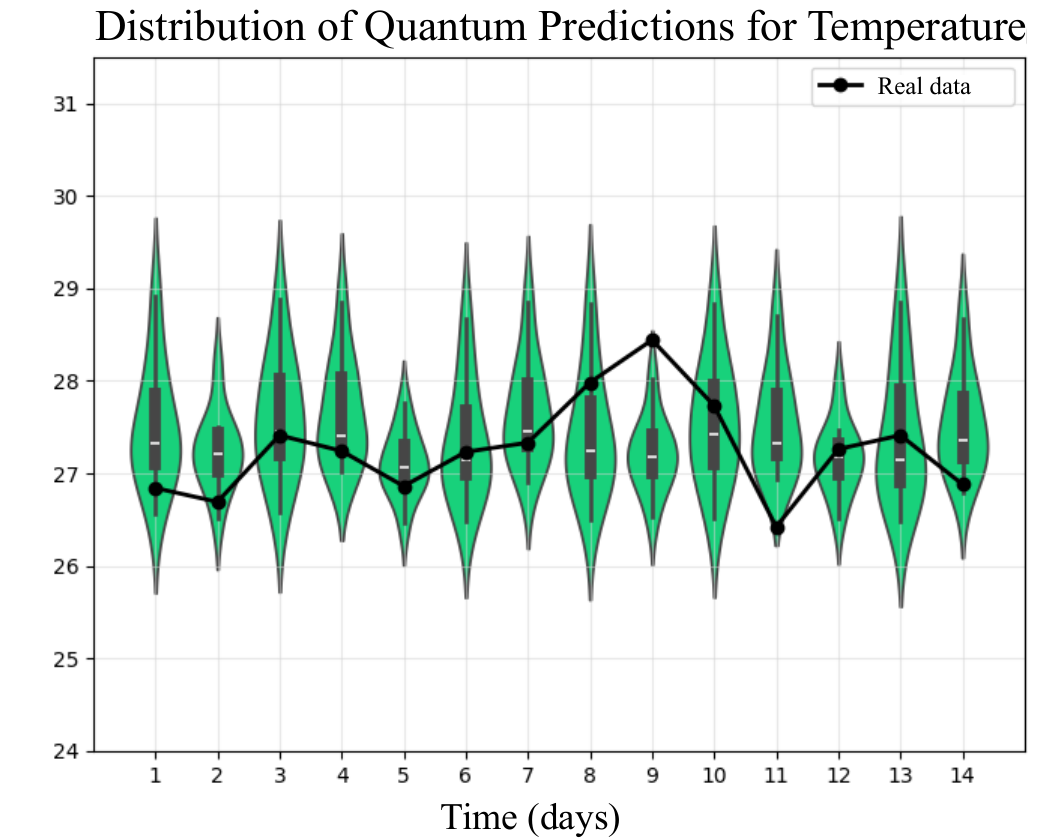}
        \caption*{Exp 2 – 1 layer}
    \end{minipage}

    \vspace{1ex}

    \begin{minipage}{0.48\textwidth}
        \centering
        \includegraphics[width=0.9\linewidth]{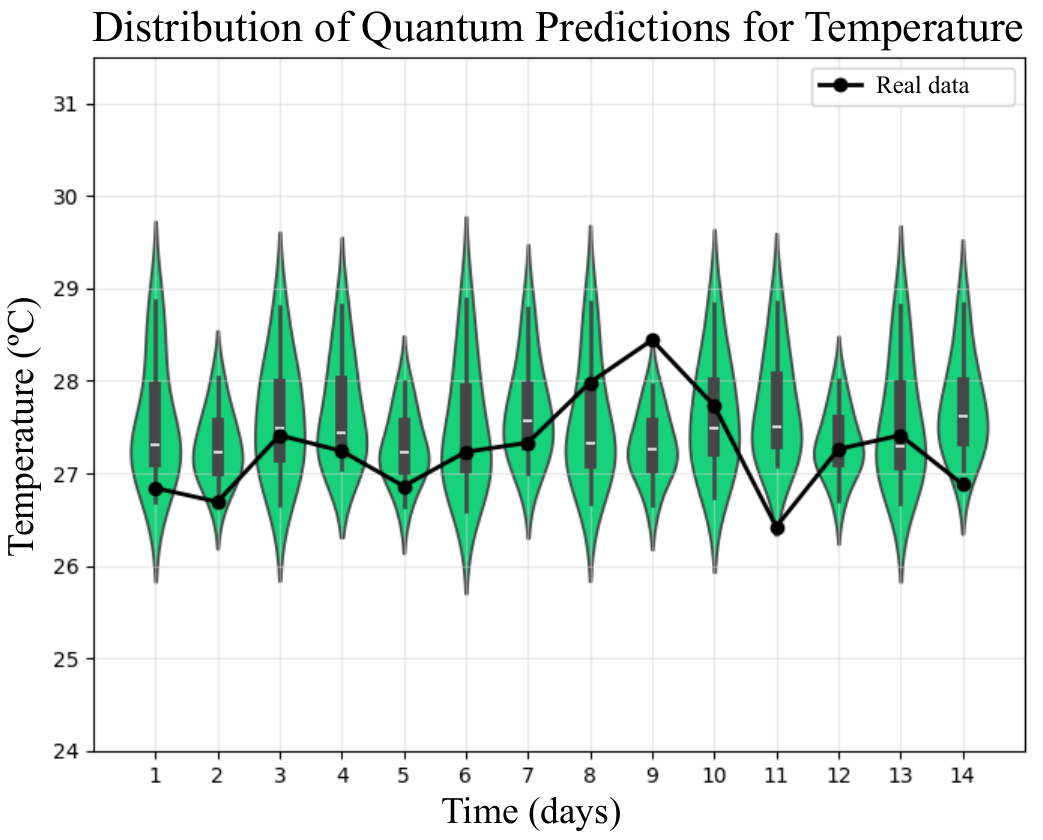}
        \caption*{Exp 1 – 3 layers}
    \end{minipage}
    \hfill
    \begin{minipage}{0.48\textwidth}
        \centering
        \includegraphics[width=0.9\linewidth]{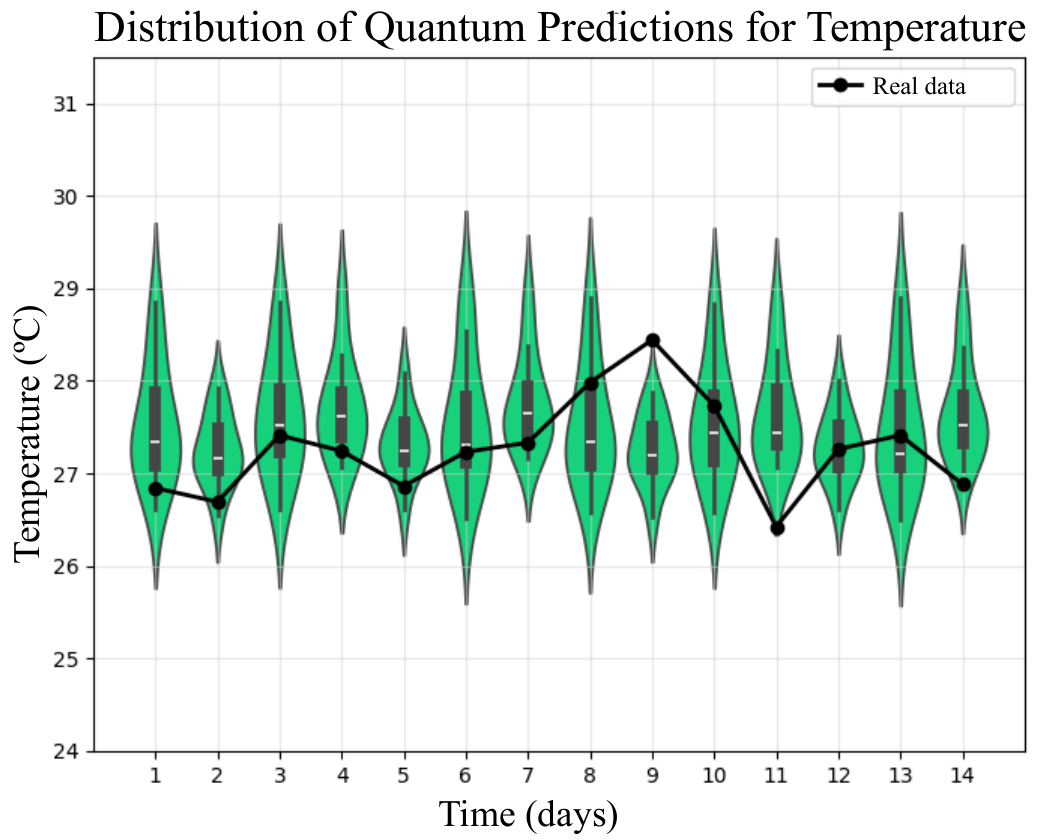}
        \caption*{Exp 2 – 3 layers}
    \end{minipage}

    \vspace{1ex}

    \begin{minipage}{0.48\textwidth}
        \centering
        \includegraphics[width=0.9\linewidth]{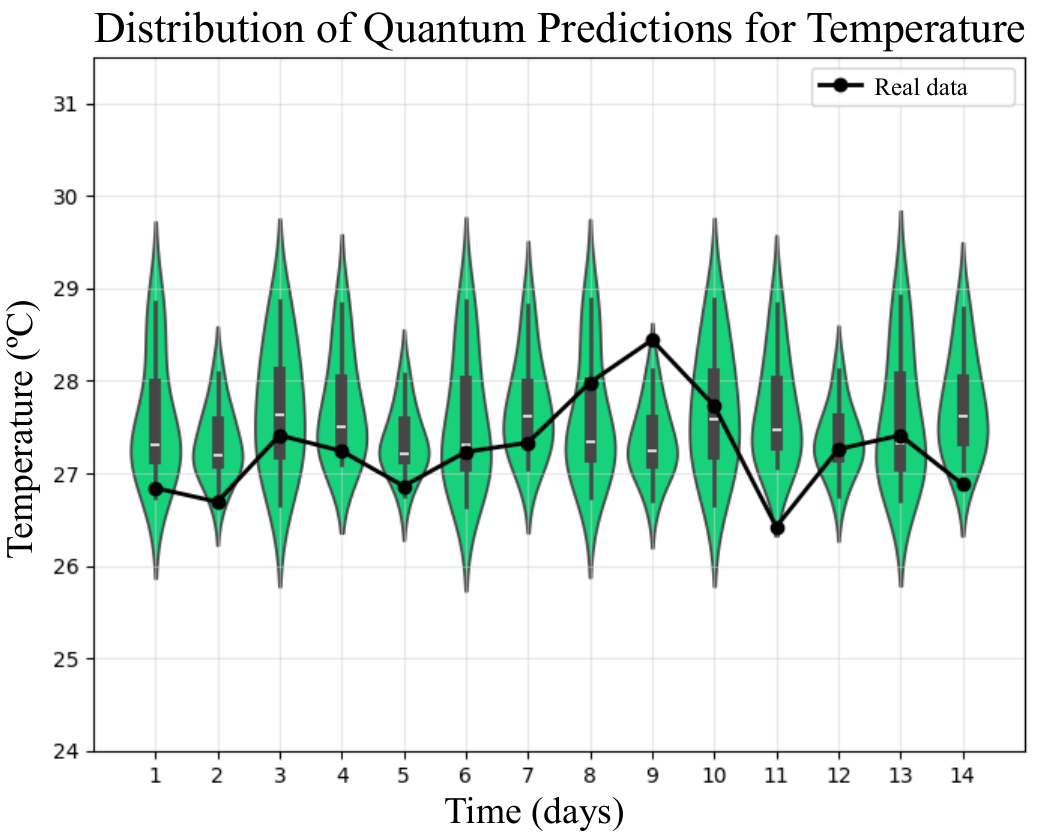}
        \caption*{Exp 1 – 5 layers}
    \end{minipage}
    \hfill
    \begin{minipage}{0.48\textwidth}
        \centering
        \includegraphics[width=0.9\linewidth]{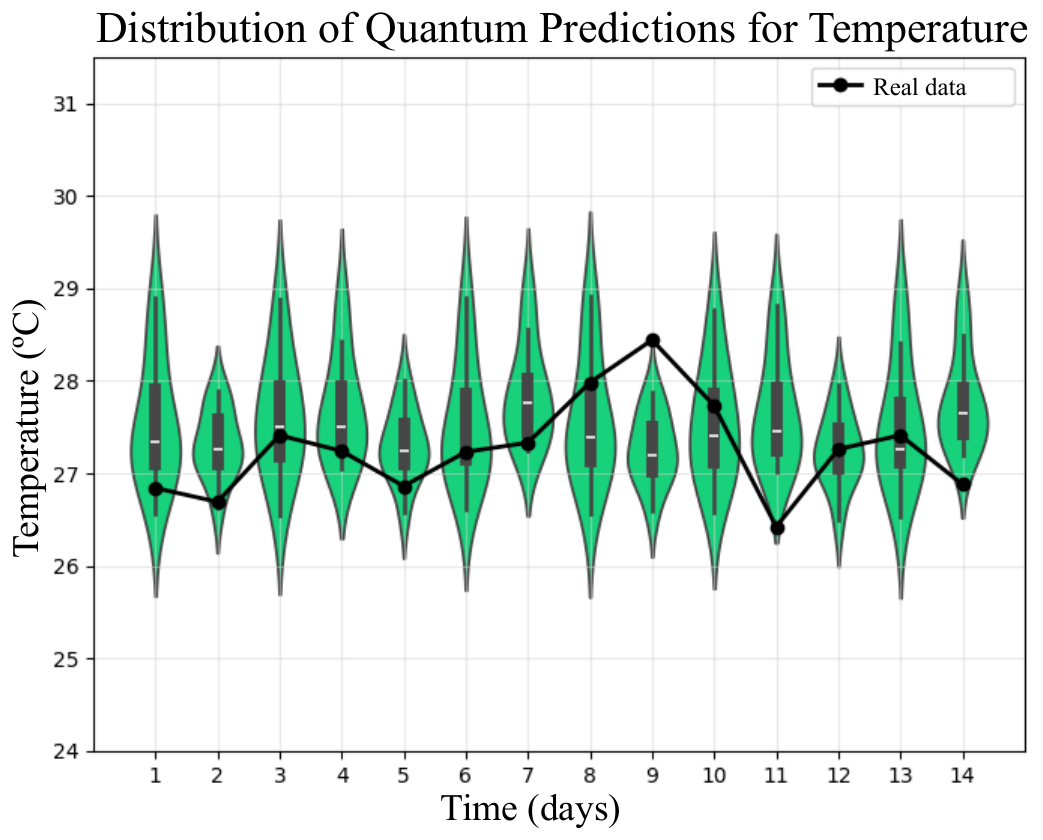}
        \caption*{Exp 2 – 5 layers}
    \end{minipage}

    \caption{ \justifying~ Distribution of quantum temperature forecasts for 14 days ahead, averaged over 10 runs per experiment.}
    \label{fig:qdist_prevtemperatura}
\end{figure*}

The second hyperparameter defined was the number of qubits in the QNN. Since we selected five climate variables along with the time-lagged target variable, the network architecture uses a total of six qubits, each encoding one feature. Table \ref{tab:config_temp} presents the experimental setup used for the temperature forecasting tasks. Results were obtained from 10 runs of each experiment for each configuration of variational layers, as previously described. Forecast distributions are shown in violin plots in Figure \ref{fig:qdist_prevtemperatura}.


\begin{table}[h!] 
\centering
\caption{ \justifying~  Experimental environment configuration for temperature forecasting.}
\label{tab:config_temp}
\begin{tabular}{p{2cm}p{3.5cm}p{3cm}}
\toprule
\textbf{Element} & \textbf{Attribute} & \textbf{Description} \\
\midrule
Training set     & 352   & Number of data points used to train the model \\
Test set         & 14    & Number of data points used to test the model \\
Features         & 6     & Number of climate variables used as input to the model \\
Qubits           & 6     & Number of qubits used to process the model \\
Epochs           & 30    & Number of complete passes through the training set \\
Learning rate    & 0.1   & Learning rate used during training \\
Validation split & 0.1   & Portion of training set used for validation \\
Batch size       & 10    & Number of samples processed before parameter updates \\
Hardware         & Nvidia Tesla V100 GPU (Kuatomu) & Hardware used to run the experiments \\
Software tools   & Pennylane, TensorFlow, Keras, Pandas, Scikit-learn, NumPy, Matplotlib & Tools used to implement the classical and quantum models \\
\bottomrule
\end{tabular}
\end{table}

\begin{widetext}

\begin{figure}[!t]
    \centering
    \begin{minipage}{0.48\textwidth}
        \centering
        \includegraphics[width=0.9\linewidth]{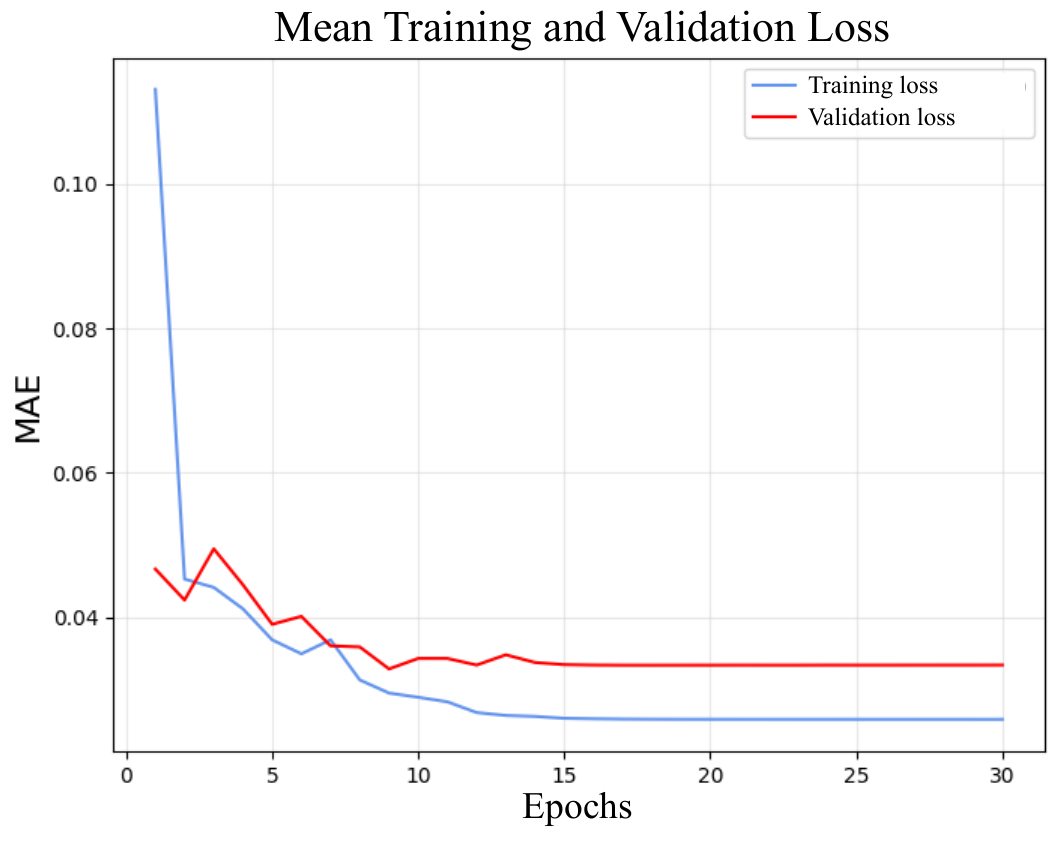}
        \caption*{(a) Experiment 1 – 1 layer.}
        \label{fig:q_losstemp1}
    \end{minipage}
    \hfill
    \begin{minipage}{0.48\textwidth}
        \centering
        \includegraphics[width=0.9\linewidth]{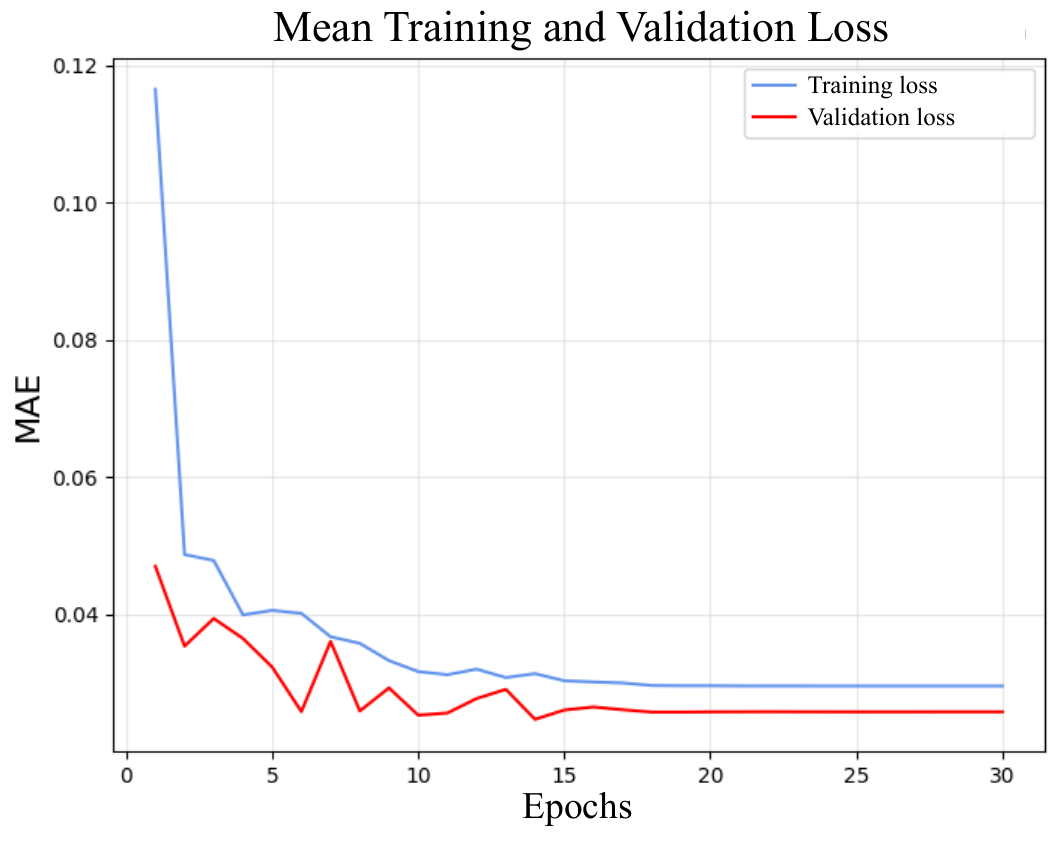}
        \caption*{(b) Experiment 2 – 1 layer.}
        \label{fig:q2_losstemp1}
    \end{minipage}

    \vspace{1ex}

    \begin{minipage}{0.48\textwidth}
        \centering
        \includegraphics[width=0.9\linewidth]{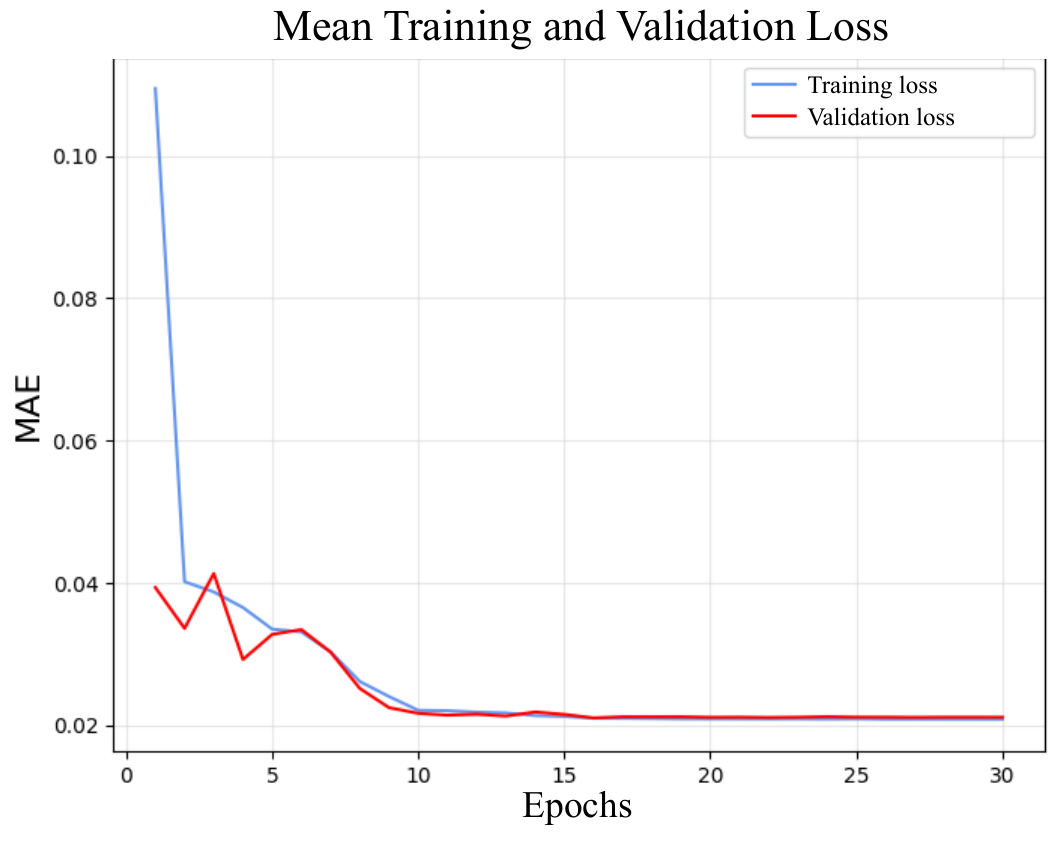}
        \caption*{(c) Experiment 1 – 3 layers.}
        \label{fig:q_losstemp3}
    \end{minipage}
    \hfill
    \begin{minipage}{0.48\textwidth}
        \centering
        \includegraphics[width=0.9\linewidth]{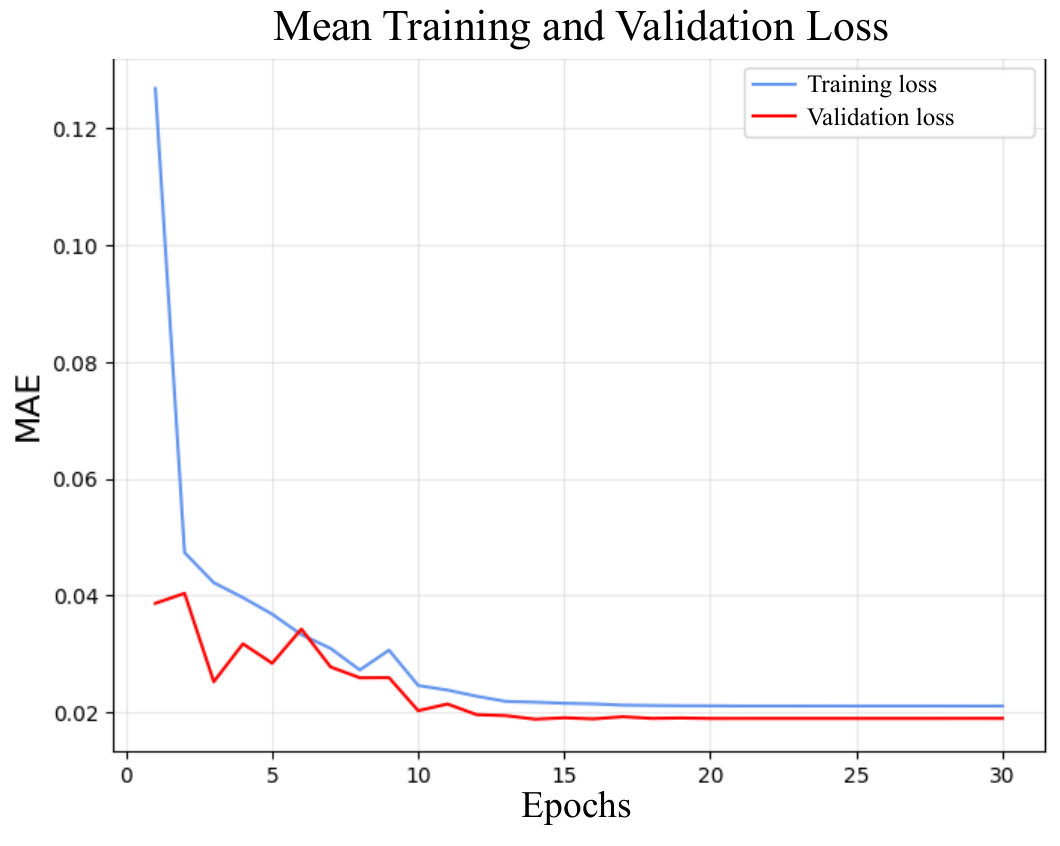}
        \caption*{(d) Experiment 2 – 3 layers.}
        \label{fig:q2_losstemp3}
    \end{minipage}

    \vspace{1ex}

    \begin{minipage}{0.48\textwidth}
        \centering
        \includegraphics[width=0.9\linewidth]{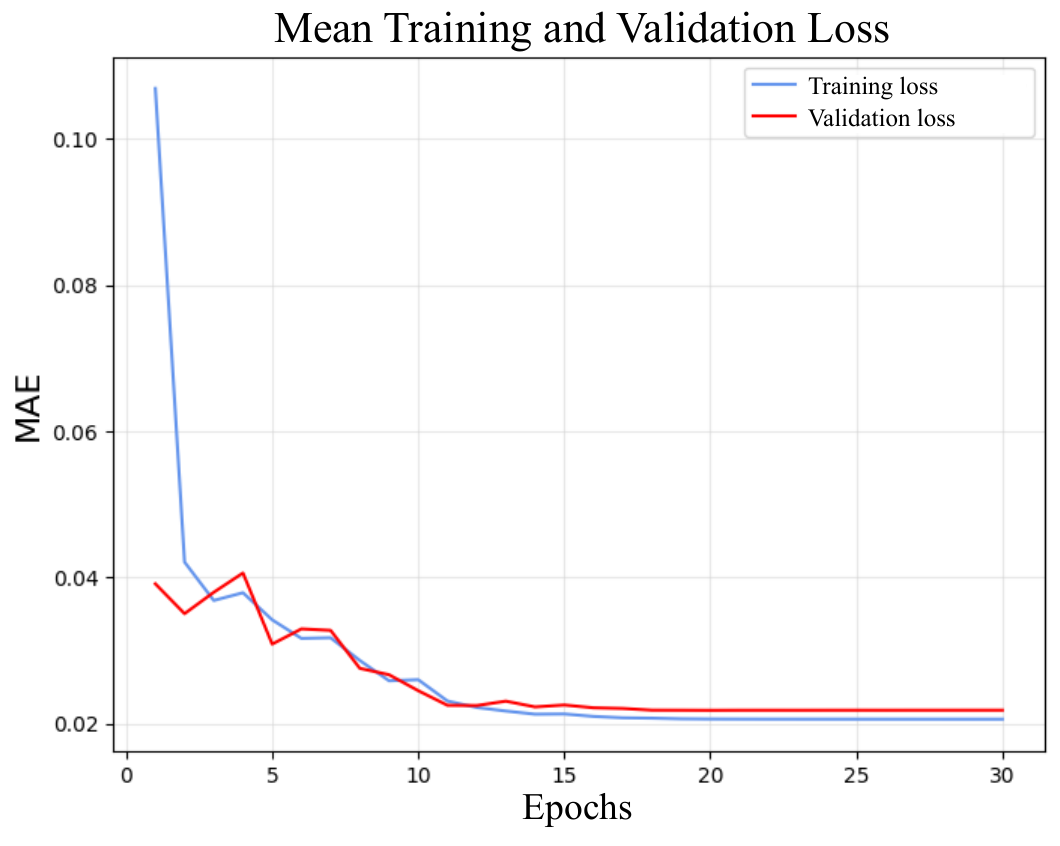}
        \caption*{(e) Experiment 1 – 5 layers.}
        \label{fig:q_losstemp5}
    \end{minipage}
    \hfill
    \begin{minipage}{0.48\textwidth}
        \centering
        \includegraphics[width=0.9\linewidth]{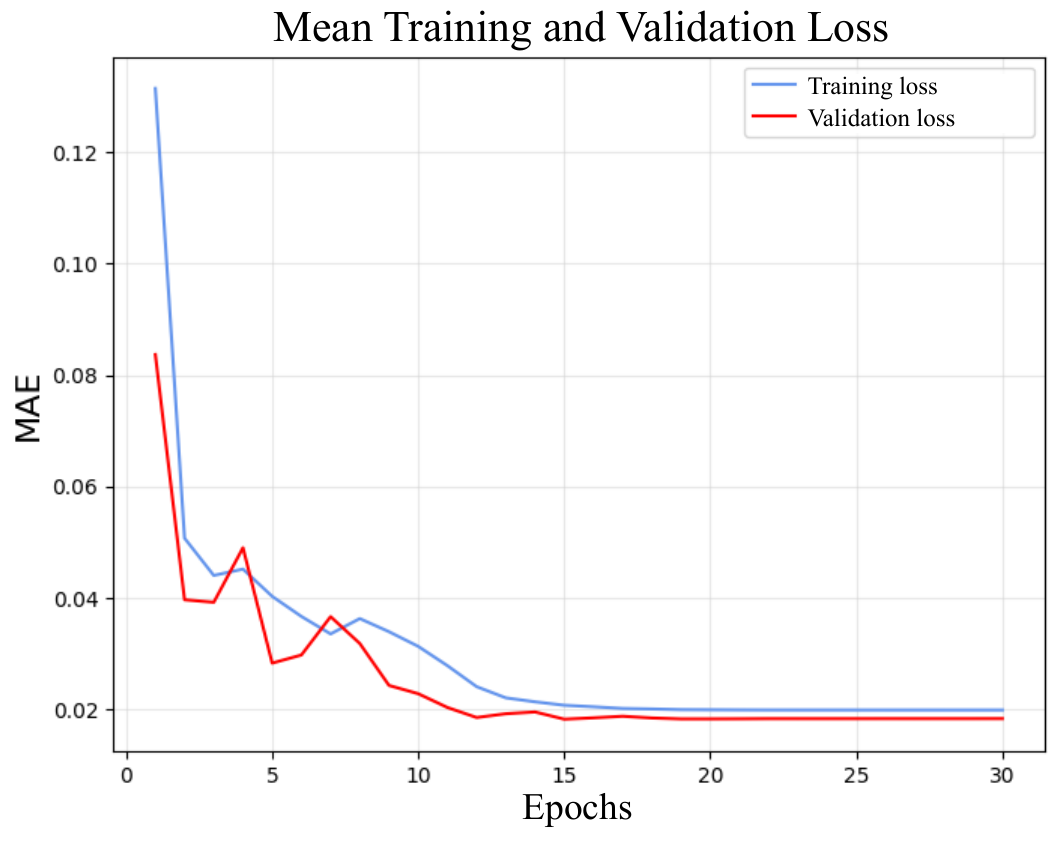}
        \caption*{(f) Experiment 2 – 5 layers.}
        \label{fig:q2_losstemp5}
    \end{minipage}

    \caption{ \justifying~Average training loss (blue) and validation loss (red) values for quantum temperature forecasting 14 days ahead, calculated as the average over 10 runs per experiment. A total of 30 epochs was used. Subfigures (a), (c), and (e) show the results of Experiment 1 using 1, 3, and 5 variational layers, respectively. Subfigures (b), (d), and (f) show the results of Experiment 2 using 1, 3, and 5 variational layers, respectively.}
    \label{fig:q_losstemperatura}
\end{figure}

\end{widetext}

Figure \ref{fig:q_losstemperatura} displays the average values of the training and validation loss functions, which reflect training efficiency and generalization ability. Finally, Figure \ref{fig:metricas_temp} compares the MAEs associated with each of the six QNN architectures, along with the classical baseline. \\

\begin{figure}[!h]
    \centering
    \includegraphics[width=0.9\linewidth]{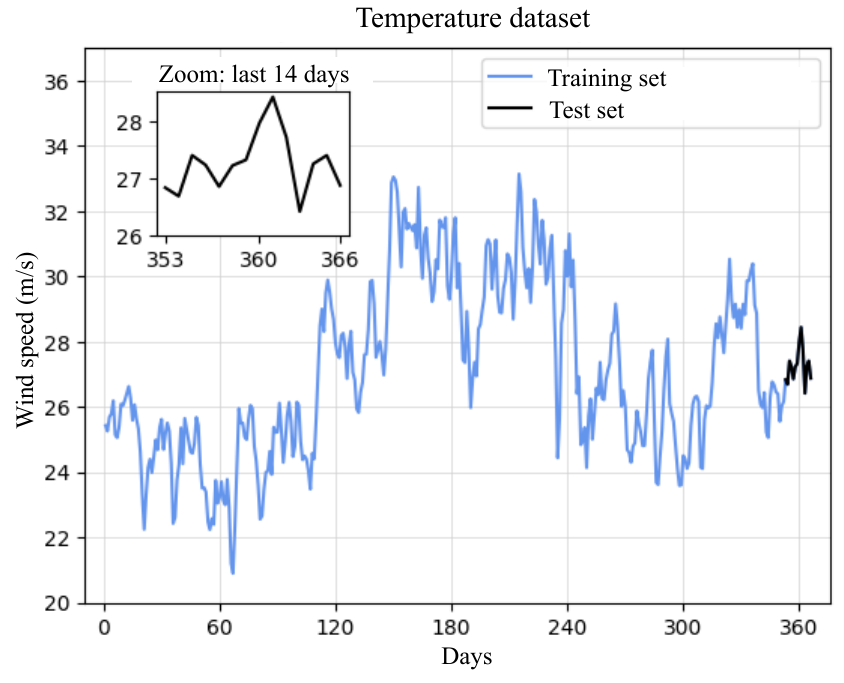}
    \caption{ \justifying~Subdivision of the dataset into training set (blue) and test set (black).}
    \label{fig:conjuntostemp}
\end{figure}


\subsubsection{Classical Model} \label{subsec1_c}

\begin{figure}[h!]
    \centering
    \includegraphics[width=0.9\linewidth]{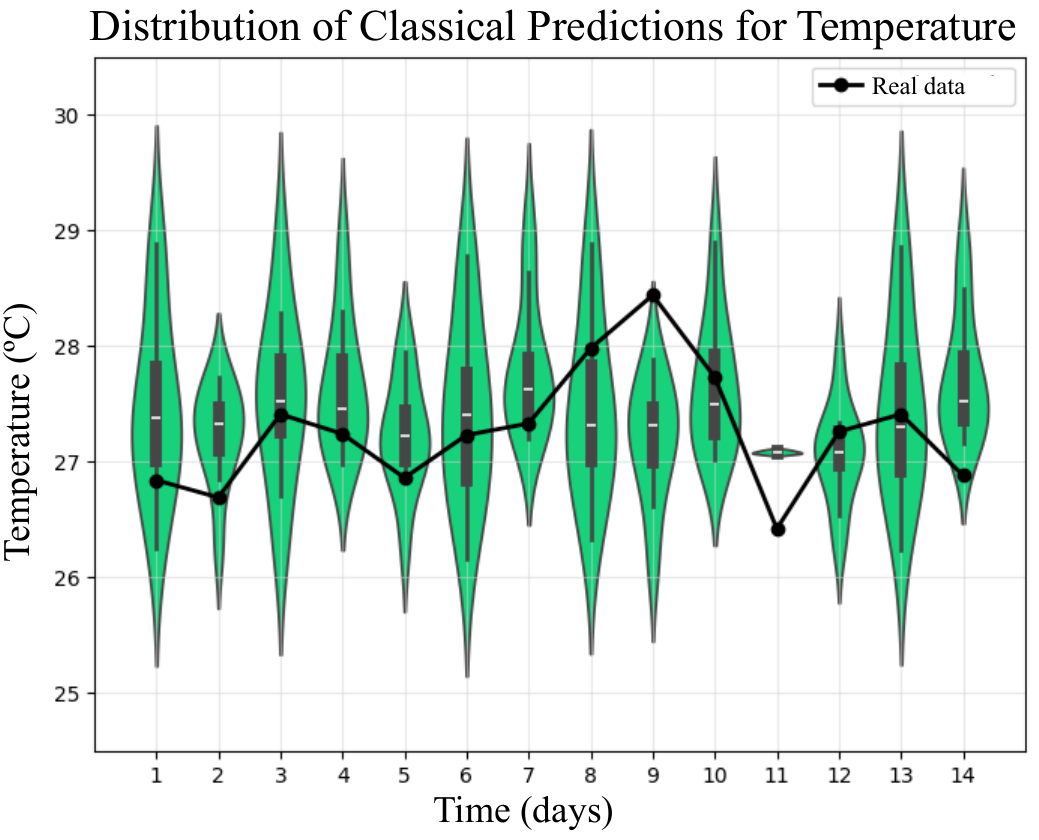}
    \caption{ \justifying~Classical temperature forecasting for 14 days ahead, calculated as the average over 10 runs.}
    \label{fig:prevclassicatemp_dist}
\end{figure}

\begin{figure}[h!]
    \centering
    \includegraphics[width=0.9\linewidth]{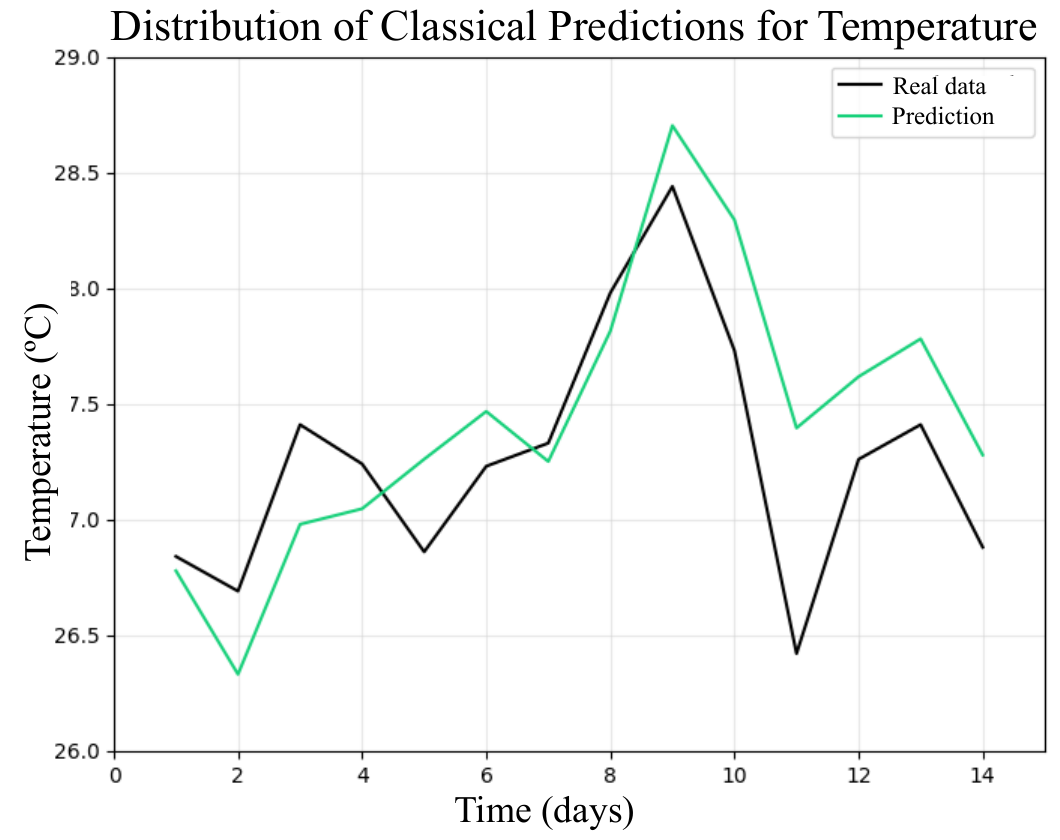}
    \caption{ \justifying~Distribution of classical temperature forecasting data for 14 days ahead, calculated as the average over 10 runs.}
    \label{fig:prevclassicatemp}
\end{figure}

\begin{figure}[h!]
    \centering
    \includegraphics[width=0.9\linewidth]{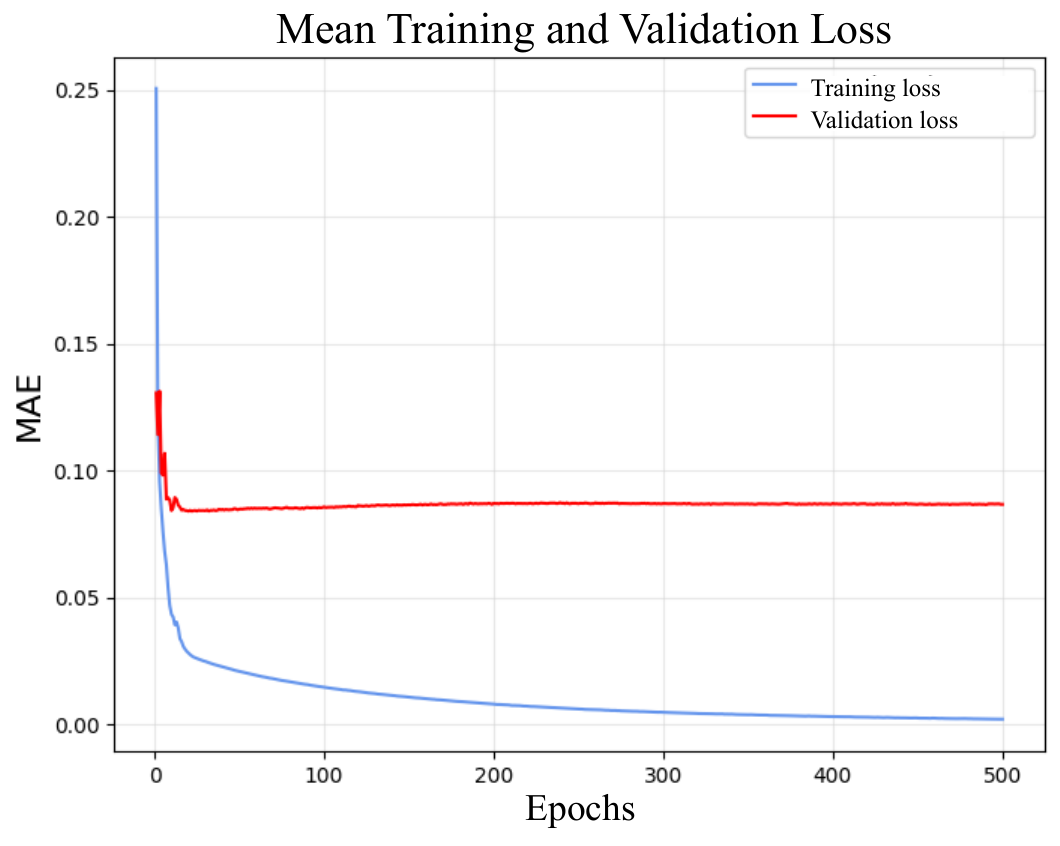}
    \caption{ \justifying~Average loss function values (blue) and validation loss values (red) for classical temperature forecasting 14 days ahead, calculated as the average over 10 runs. A total of 500 epochs was used.}
    \label{fig:lossclassicatemp}
\end{figure}

The RNN used for classical temperature forecasting was configured with the same number of \textit{features}, $validation\_split$, $batch\_size$, hardware, and training and testing proportions as the QNN, as shown in Table \ref{tab:config_temp}. However, due to its classical nature, incompatible with the use of qubits, 256 neurons were employed. Another modified hyperparameter was the initial learning rate, which was set to $lr = 0.001$. Additionally, the number of epochs was fixed at 500 epochs to analyze the behavior of the training and validation \textit{losses}. Figure \ref{fig:prevclassicatemp_dist} shows the distribution of the classical temperature forecast, followed by the graphical representation of its respective mean value in Figure \ref{fig:prevclassicatemp} and the average training and validation losses over the 500 implemented epochs, as shown in Figure \ref{fig:lossclassicatemp}.

Figure \ref{fig:prevclassicatemp_dist} displays violin plots encompassing the real data for all 14 forecast days, where a high variation in the distribution of runs is observed. Overall, the forecasts were accurate, as they fall within the normal range of the violins. Exceptions are days 2, 9, and 14, which lie in the outlier region. Additionally, the model struggled to predict the temperature on day 11, probably due to an abrupt drop compared to the previous day's value.

The mean value of the 10 algorithm repetitions is displayed in Figure \ref{fig:prevclassicavento} and shows difficulties capturing the temporal trend between days 3 and 7. Finally, Figure \ref{fig:lossclassicavento} revealed convergence of the training loss at 0.01 and validation loss at 0.08. It is observed that early in training, the model showed mild generalization difficulties, as indicated by the initial instability of the red curve; however, the corresponding increase in training loss stabilized the generalization capability at a lower loss.

\subsubsection*{Comparison between predictive models}

As demonstrated by the violin plots in this section, the quantum forecast given by the QNN generally presents lower temperature variation, indicating more consistent predictions over time. On the other hand, the high variation associated with the classical forecast given by the RNN indicates a greater sensitivity of this model, possibly due to its ability to capture more uncertainties from real data. The accuracy of the models is presented in Table \ref{tab:metricas_temp}, showing better results especially for the quantum experiments 1, with 3 variational layers, and 2 with 1 and 3 variational layers.

In particular, on the 11th forecast day, a sharp temperature drop occurred, allowing for an evaluation of the models’ ability to adjust to extreme events. On this day, the RNN presented a scatter plot that was highly concentrated at a temperature higher than the actual one, demonstrating difficulty in predicting abrupt events. Conversely, the violin plot generated by the QNN showed a distribution that encompassed the real data at its lower extreme, outside the whisker region, indicating an outlier. Thus, the QNN demonstrated better adaptation to abrupt variations and lower dispersion of predictions than the RNN for the 14-day temperature forecast.

\begin{figure}[h!]
    \centering
    \includegraphics[scale=0.4]{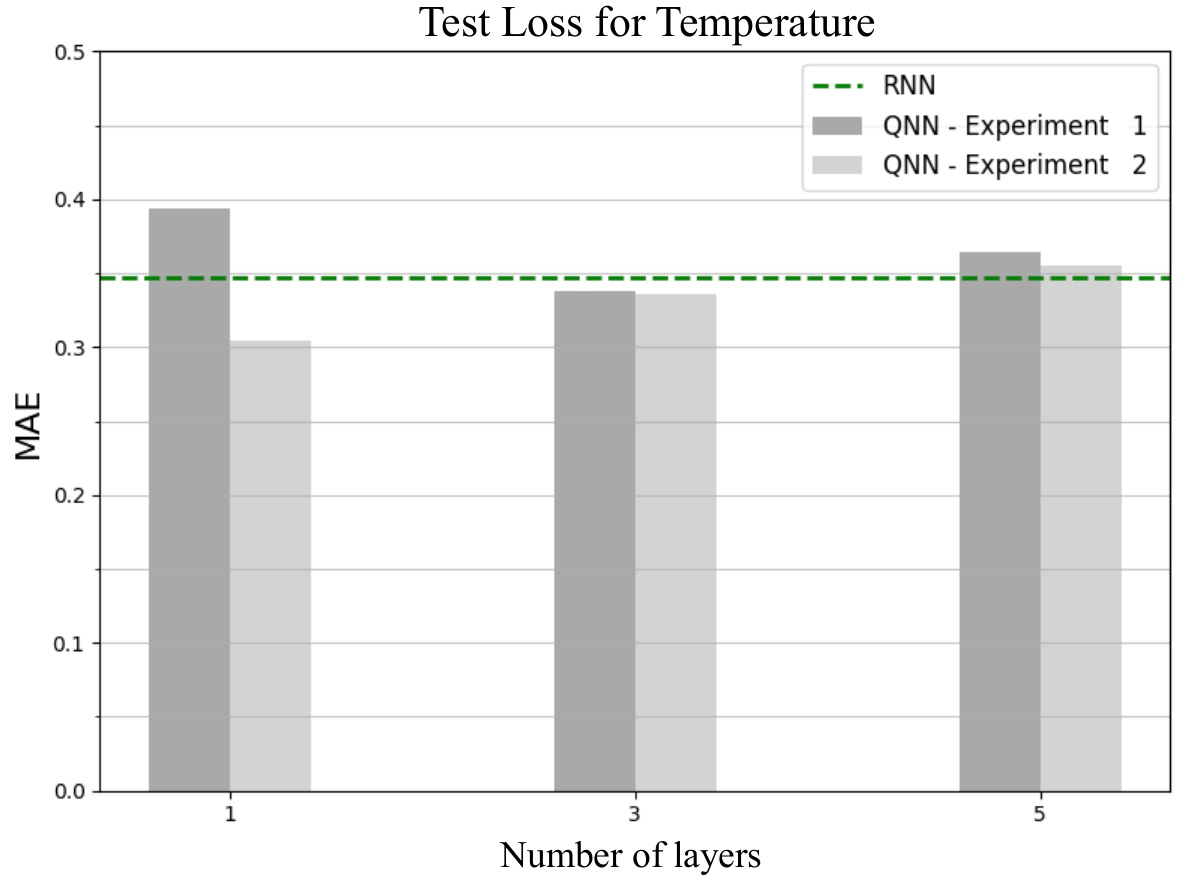}
    \caption{ \justifying~Mean absolute error associated with the loss functions on the test set for temperature forecasting. The green dashed line indicates the MAE achieved by the RNN; the dark gray bars represent the MAEs associated with QNN experiment 1, and the light gray bars represent the MAEs from QNN experiment 2. The analysis was performed for 1, 3, and 5 layers.}
    \label{fig:metricas_temp}
    \vspace{0.5cm}  
\end{figure}

The MAE and accuracy metrics achieved by the predictive models for temperature forecasting are summarized below, in Table \ref{tab:metricas_temp} and Figure \ref{fig:metricas_temp}.

\begin{table}[h!] 
\centering
\caption{ \justifying~  Performance metrics (MAE and Accuracy) obtained by the predictive models QNN and RNN for temperature forecasting.}
\label{tab:metricas_temp}
\begin{tabular}{@{}llccc@{}}
\toprule
\textbf{Model} & \textbf{Experiment} & \textbf{Layer} & \textbf{MAE} & \textbf{Accuracy (\%)} \\
\midrule
\multirow{3}{*}{QNN} 
 & \multirow{3}{*}{Experiment 1} & 1 & 0.394 & 60.6 \\
 &                                & 3 & 0.338 & 66.2 \\
 &                                & 5 & 0.364 & 63.6 \\
\cmidrule(lr){2-5}
 & \multirow{3}{*}{Experiment 2} & 1 & 0.304 & 69.6 \\
 &                                & 3 & 0.336 & 66.4 \\
 &                                & 5 & 0.355 & 64.5 \\
\multirow{1}{*}{RNN} 
 & - & - & 0.357 & 65.3 \\
\bottomrule
\end{tabular}
\end{table}

As previously discussed, the MAE of 0.347 achieved by the RNN was intermediate, outperforming half of the QNN simulations. In the quantum model, Experiment 1 exhibited nonlinear behavior, reaching its optimal performance with three variational layers. In contrast, in Experiment 2, increasing the number of layers led to a progressive rise in MAE, indicating that deeper network architectures had a detrimental impact on predictive performance. The best result in this section corresponds to QNN Experiment 2 with a single variational layer, achieving a MAE of 0.304. An anticipated result based on the respective validation loss behavior.

Therefore, it is inferred that the variational layer structure employed in Experiment 2 demonstrated the most stable behavior for forecasting temperature 14 days ahead, outperforming the other configurations and achieving optimal performance with a neural network with only one layer.

\subsection{Wind prediction}\label{subsec2}

\begin{figure}[h!]
    \centering
    \includegraphics[width=0.9\linewidth]{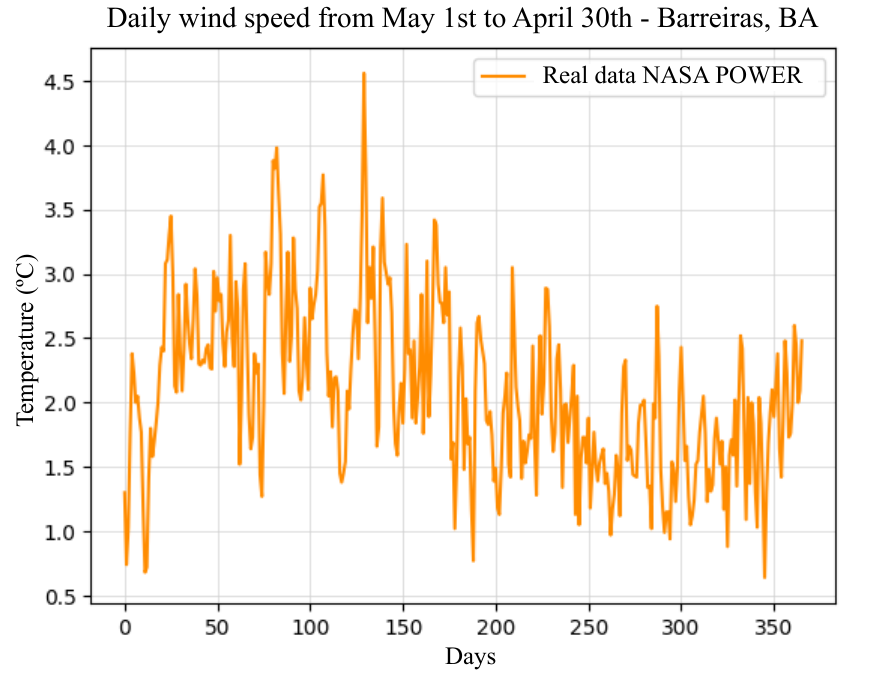}
    \caption{ \justifying~Actual daily wind speed data obtained from the NASA POWER dataset.}
    \label{fig:dadosreaisvento}
\end{figure}

\begin{figure}[h!]
    \centering
    \includegraphics[width=0.9\linewidth]{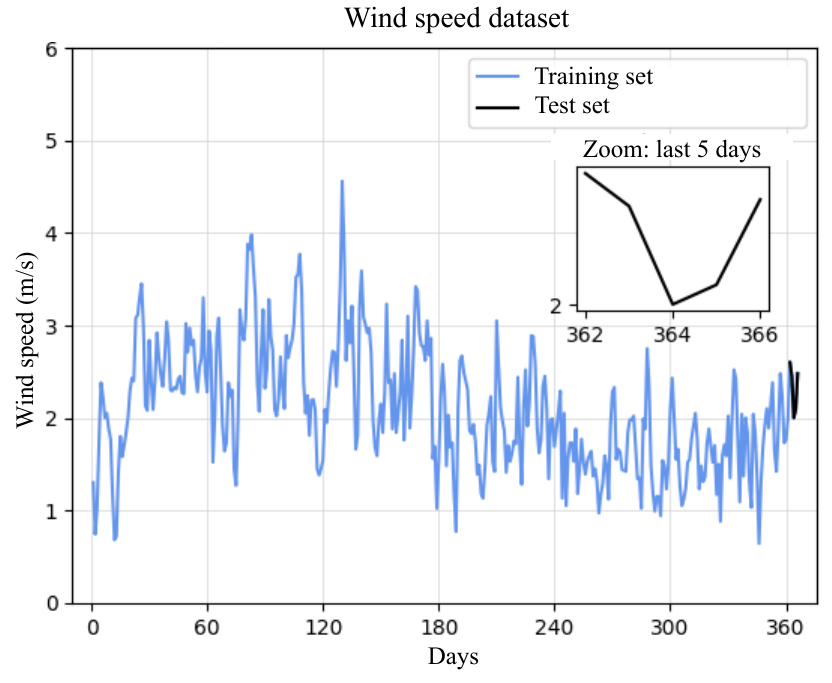}
    \caption{ \justifying~Subdivision of the daily wind speed dataset into training (blue) and testing (black) sets.}
    \label{fig:conjuntosvento}
\end{figure}

Subsequently, wind speed forecasting experiments were conducted using a train-test split based on a 6-day time lag. The prediction horizon was set to 5 days, corresponding to the size of the test set. Given that the dataset spans one year (from January 5, 2023, to April 30, 2024), the data were divided into training and testing subsets, comprising 98.7\% and 1.3\%, respectively. Figure \ref{fig:dadosreaisvento} presents the daily temperature data for the specified period and location, while Figure \ref{fig:conjuntosvento} illustrates the previously calculated test and training proportions.

\subsubsection{Quantum Model} \label{subsec2_q}

\begin{figure*}[!t]
    \centering
    \begin{minipage}{0.48\textwidth}
        \centering
        \includegraphics[width=0.9\linewidth]{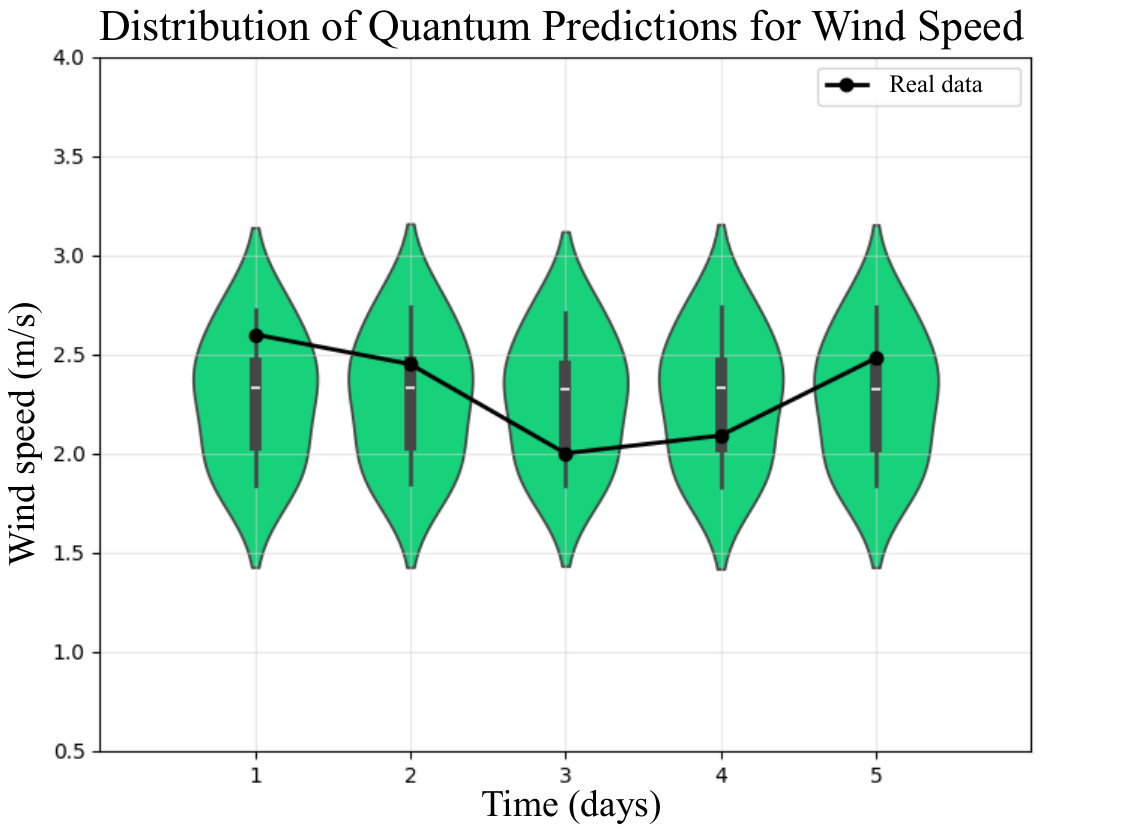}
        \caption*{(a) Experiment 1 – 1 layer.}
        \label{fig:qdist_prevvento1}
    \end{minipage}
    \hfill
    \begin{minipage}{0.48\textwidth}
        \centering
        \includegraphics[width=0.9\linewidth]{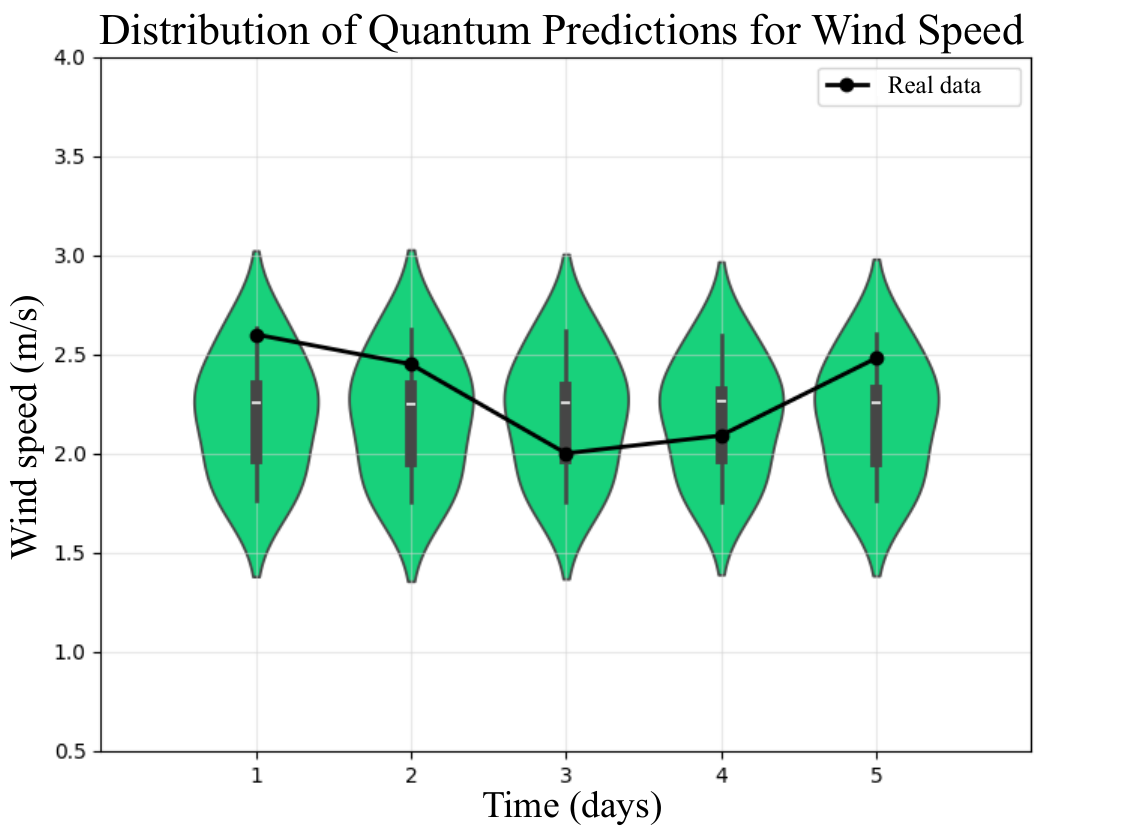}
        \caption*{(b) Experiment 2 – 1 layer.}
        \label{fig:q2dist_prevvento1}
    \end{minipage}

    \vspace{1ex}

    \begin{minipage}{0.48\textwidth}
        \centering
        \includegraphics[width=0.9\linewidth]{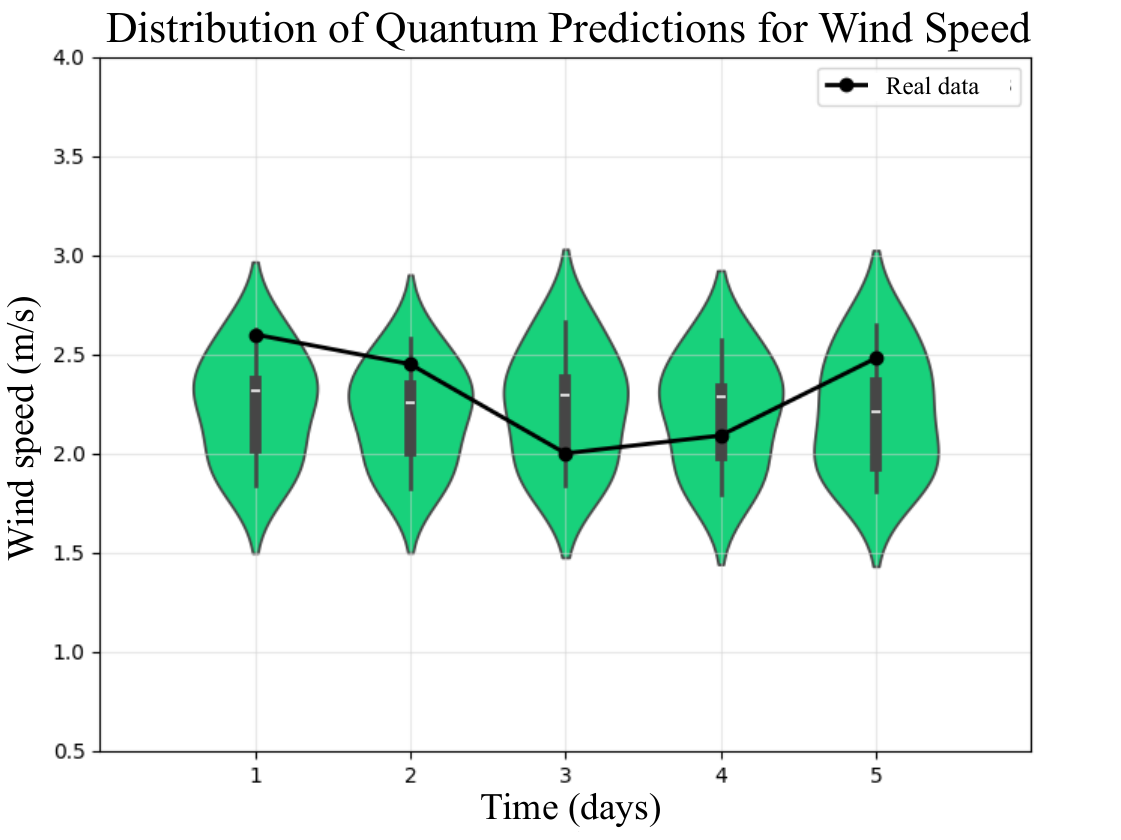}
        \caption*{(c) Experiment 1 – 3 layers.}
        \label{fig:qdist_prevvento3}
    \end{minipage}
    \hfill
    \begin{minipage}{0.48\textwidth}
        \centering
        \includegraphics[width=0.9\linewidth]{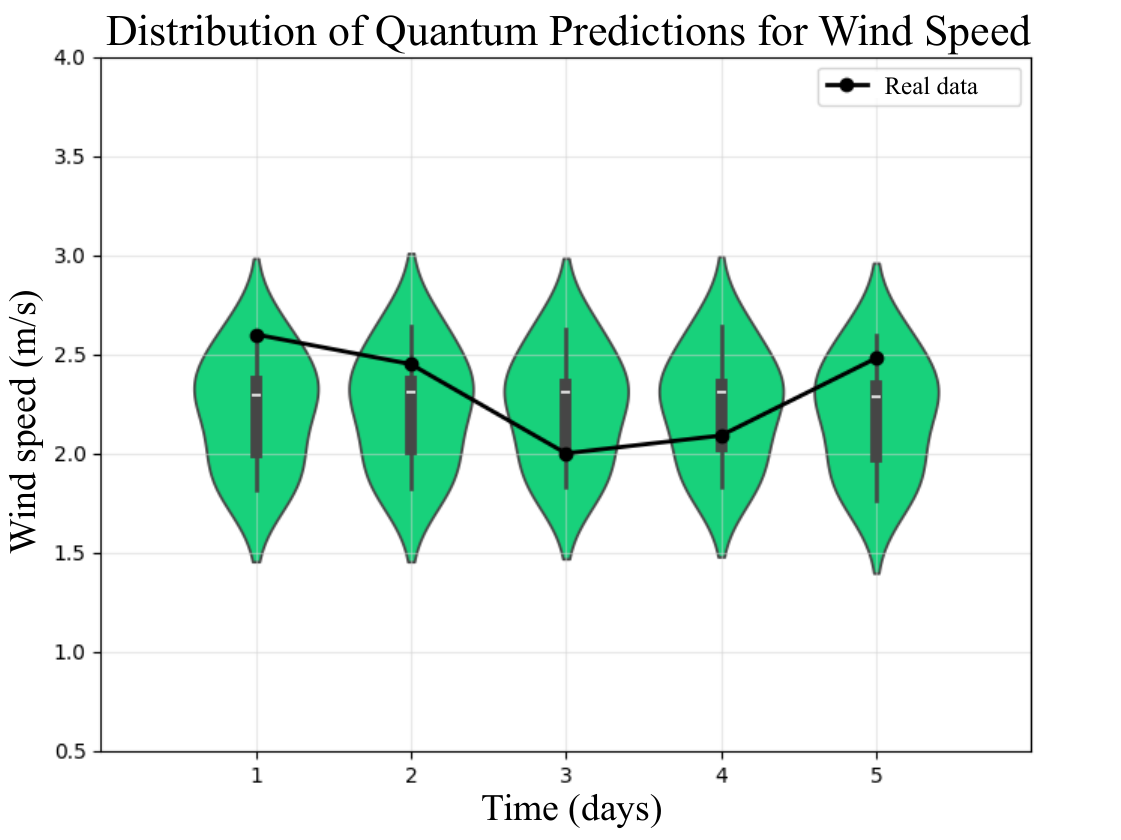}
        \caption*{(d) Experiment 2 – 3 layers.}
        \label{fig:q2dist_prevvento3}
    \end{minipage}

    \vspace{1ex}

    \begin{minipage}{0.48\textwidth}
        \centering
        \includegraphics[width=0.9\linewidth]{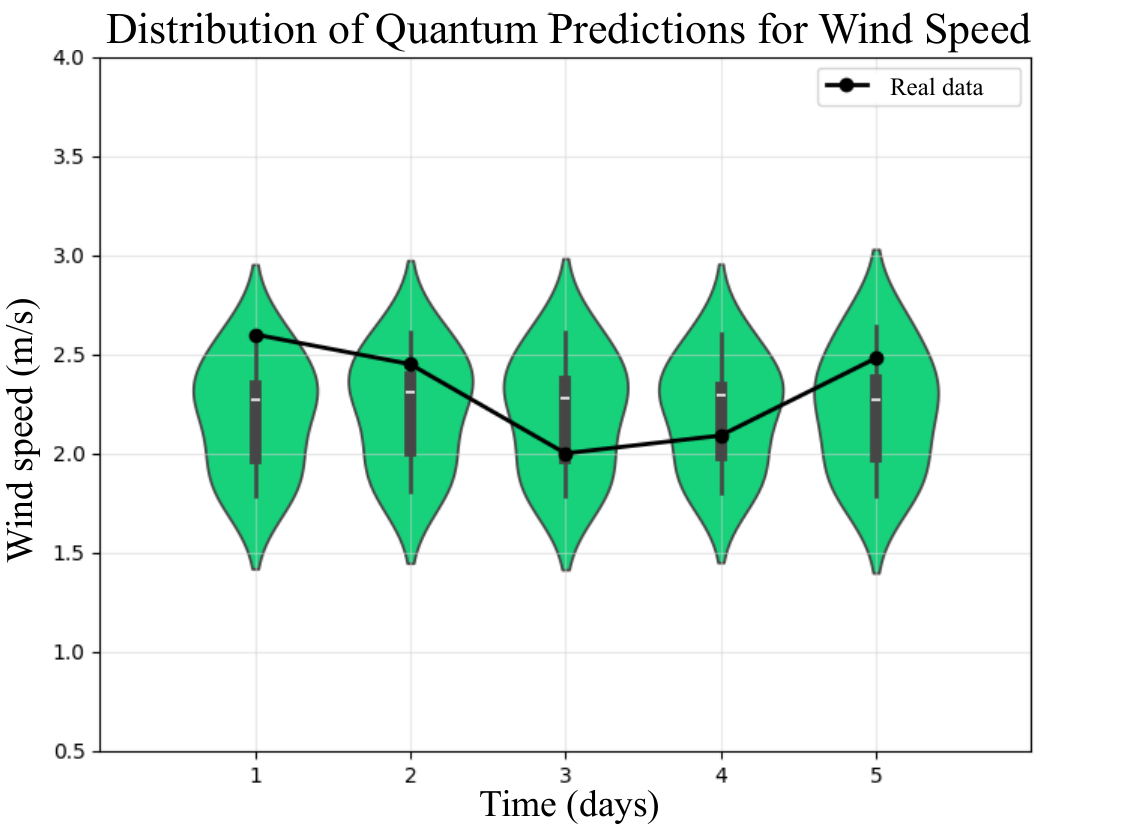}
        \caption*{(e) Experiment 1 – 5 layers.}
        \label{fig:qdist_prevvento5}
    \end{minipage}
    \hfill
    \begin{minipage}{0.48\textwidth}
        \centering
        \includegraphics[width=0.9\linewidth]{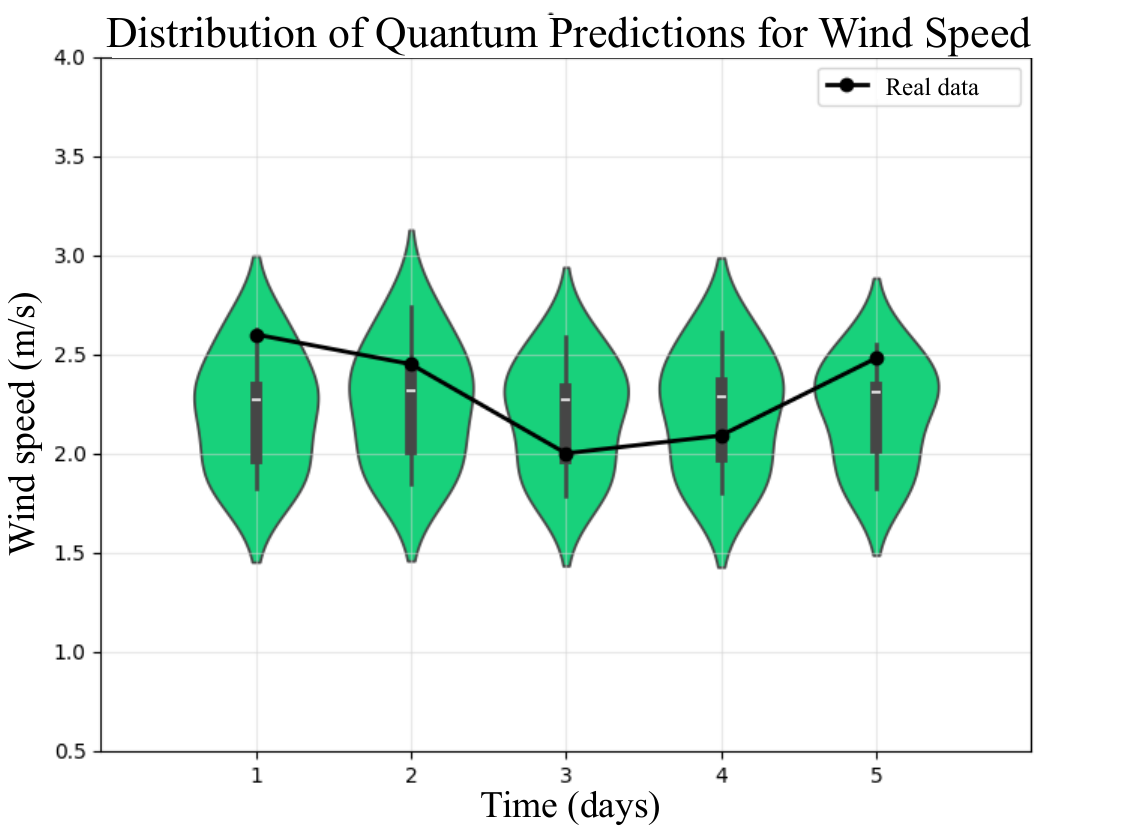}
        \caption*{(f) Experiment 2 – 5 layers.}
        \label{fig:q2dist_prevvento5}
    \end{minipage}

    \caption{ \justifying~Distribution of 5-day-ahead wind forecast data based on the average of 10 runs per experiment. Subfigures (a), (c), and (e) show the results of Experiment 1 with 1, 3, and 5 variational layers, respectively. Subfigures (b), (d), and (f) show the results of Experiment 2 under the same conditions.}
    \label{fig:qdist_prevvento}
\end{figure*}

The second hyperparameter defined was the number of qubits in the QNN, as seven climate variables were previously selected, along with the time-lagged target variable; the network architecture used a total of eight qubits, each encoding one feature. Table \ref{tab:config_vento} summarizes the experimental setup used for the wind speed forecasting tasks. Results were obtained from 10 runs for each configuration of variational layers, as previously described. Forecast distributions are presented in violin plot format in Figure \ref{fig:qdist_prevvento}.

\begin{figure*}[!t]
    \centering
    \begin{minipage}{0.48\textwidth}
        \centering
        \includegraphics[width=0.9\linewidth]{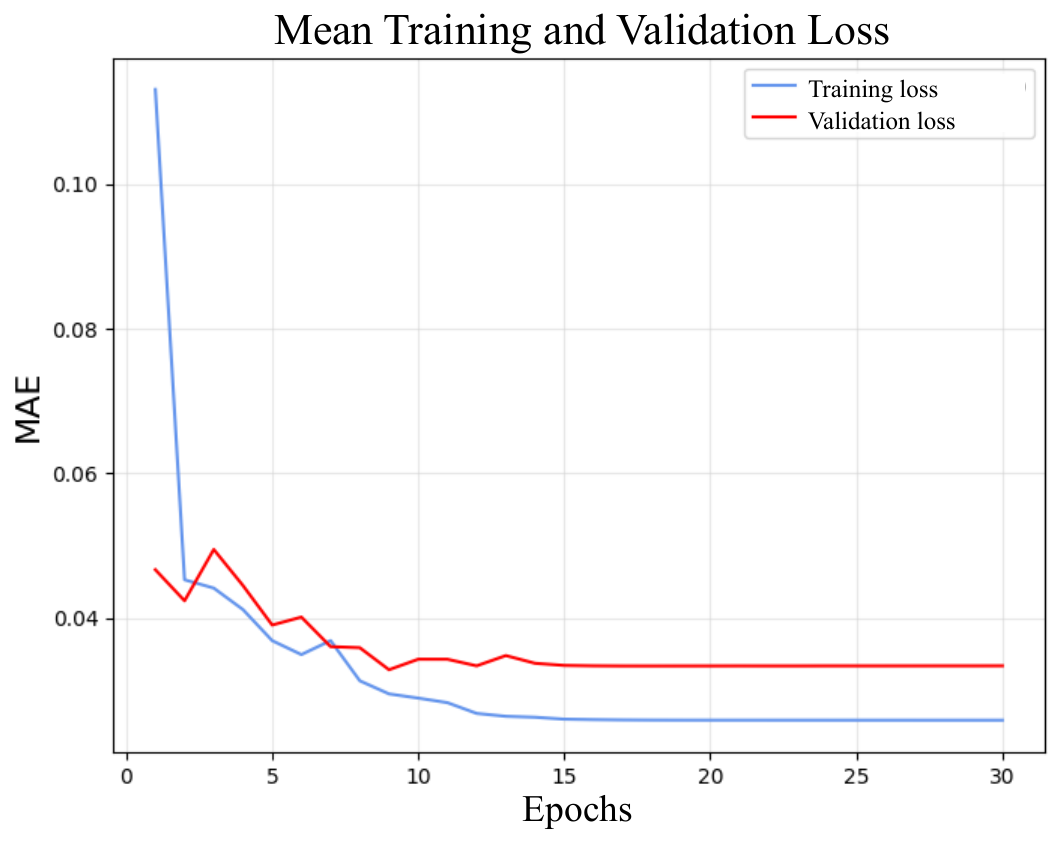}
        \caption*{(a) Experiment 1 – 1 layer.}
        \label{fig:q_lossvento1}
    \end{minipage}
    \hfill
    \begin{minipage}{0.48\textwidth}
        \centering
        \includegraphics[width=0.9\linewidth]{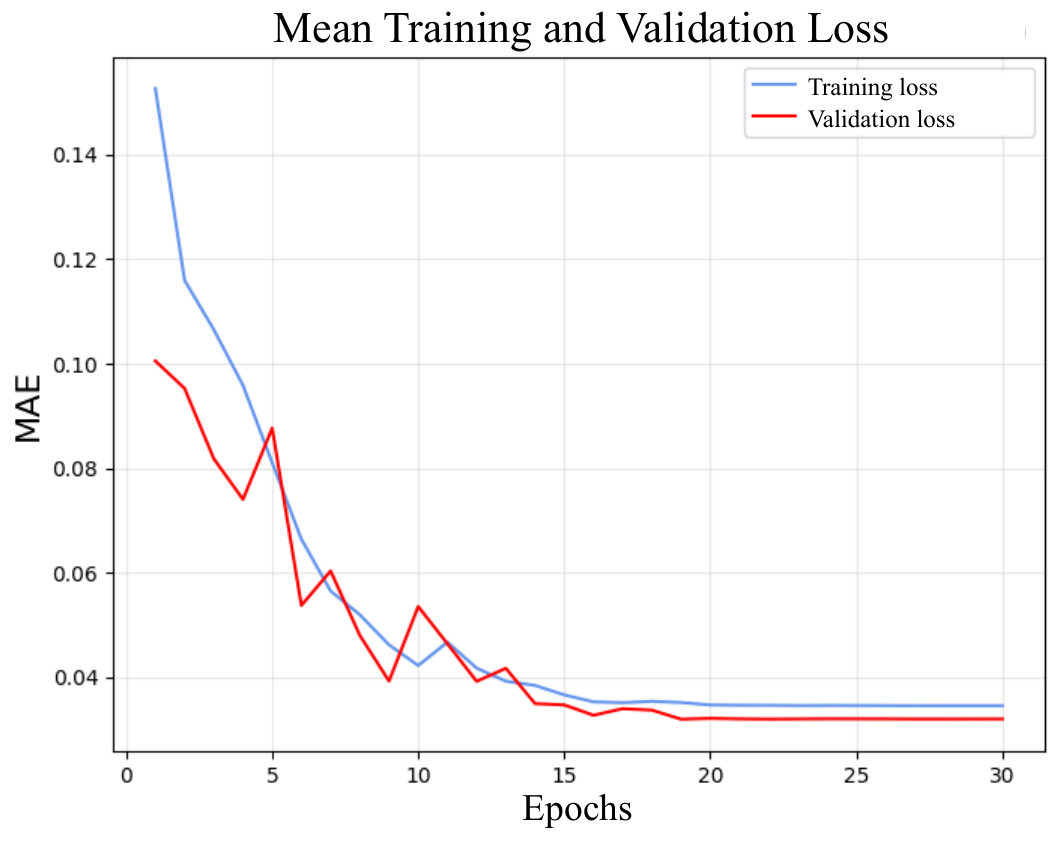}
        \caption*{(b) Experiment 2 – 1 layer.}
        \label{fig:q2_lossvento1}
    \end{minipage}

    \vspace{1ex}

    \begin{minipage}{0.48\textwidth}
        \centering
        \includegraphics[width=0.9\linewidth]{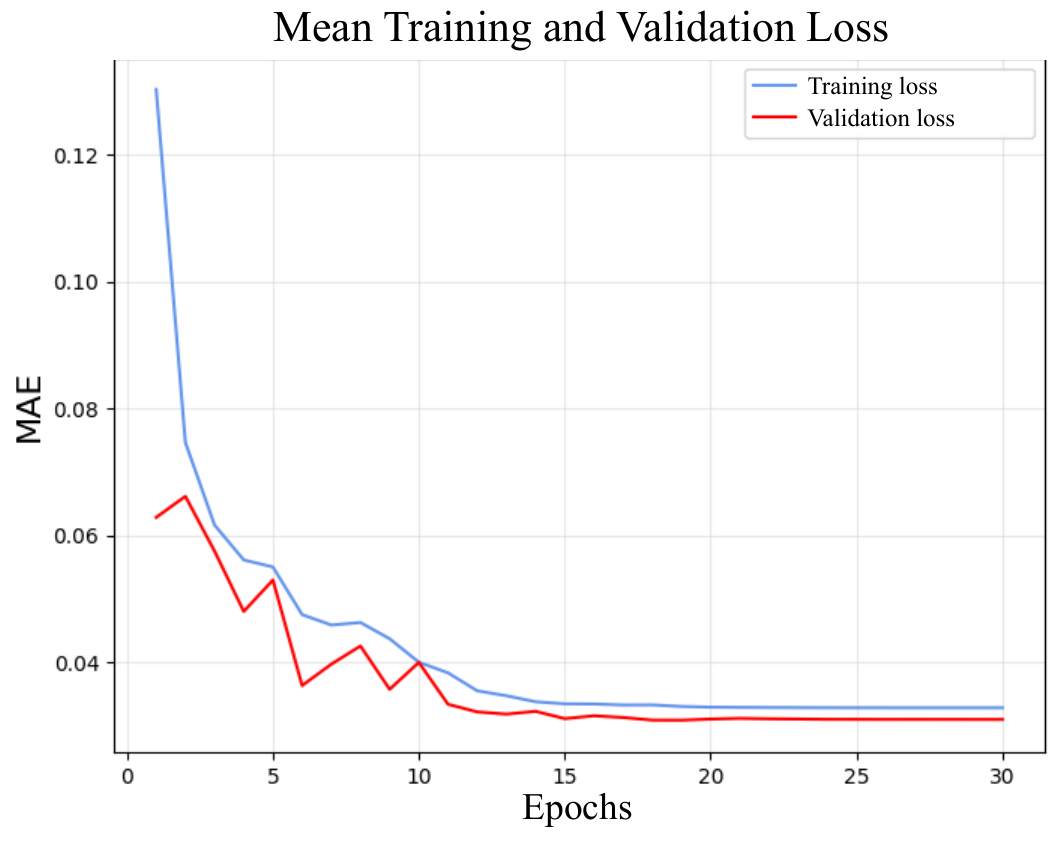}
        \caption*{(c) Experiment 1 – 3 layers.}
        \label{fig:q_lossvento3}
    \end{minipage}
    \hfill
    \begin{minipage}{0.48\textwidth}
        \centering
        \includegraphics[width=0.9\linewidth]{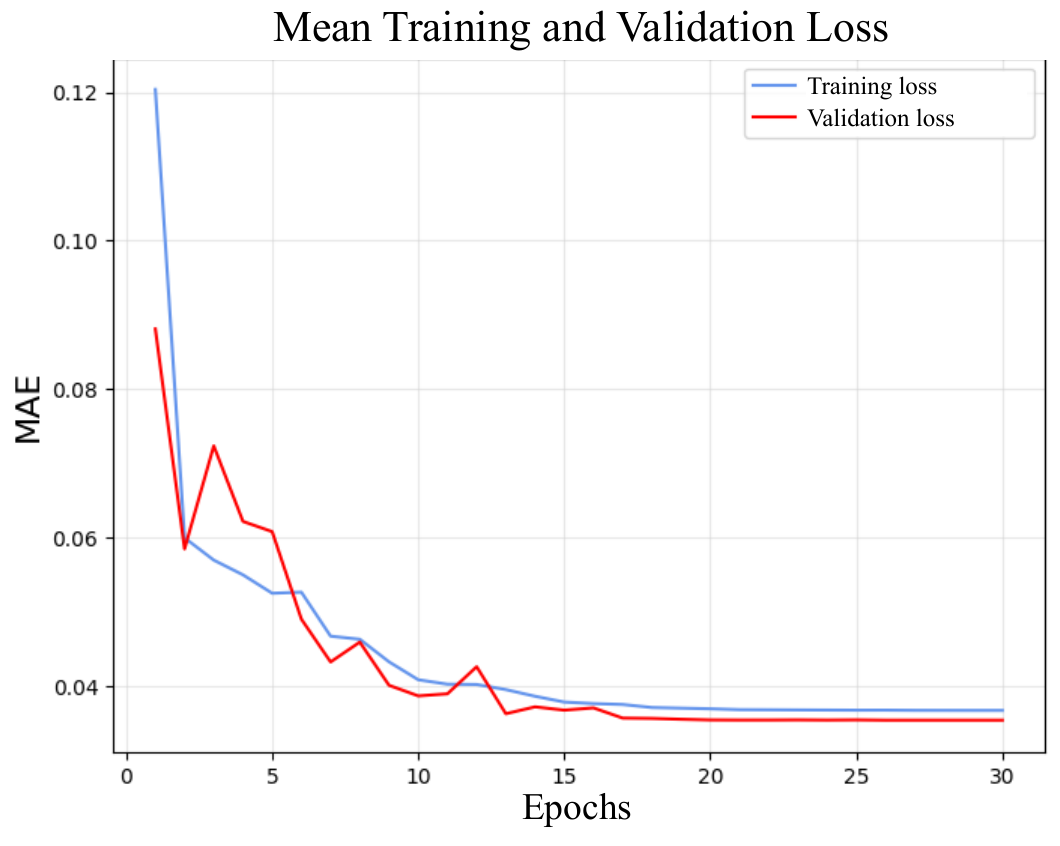}
        \caption*{(d) Experiment 2 – 3 layers.}
        \label{fig:q2_lossvento3}
    \end{minipage}

    \vspace{1ex}

    \begin{minipage}{0.48\textwidth}
        \centering
        \includegraphics[width=0.9\linewidth]{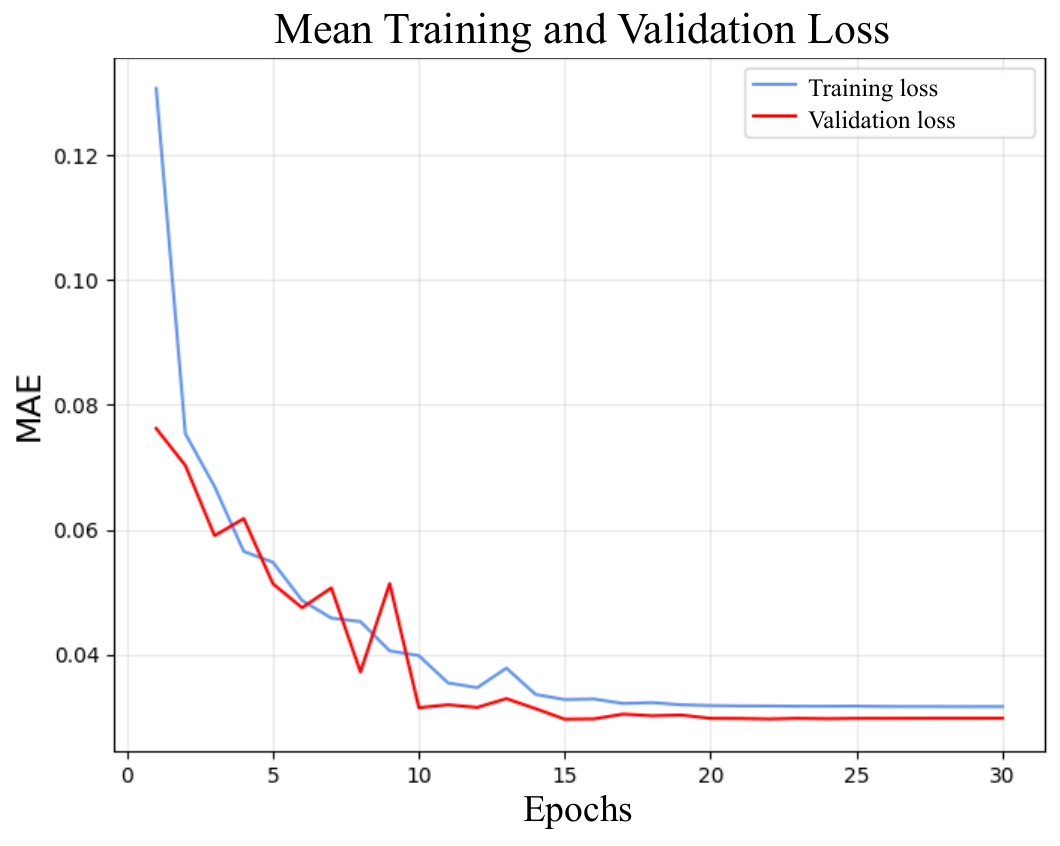}
        \caption*{(e) Experiment 1 – 5 layers.}
        \label{fig:q_lossvento5}
    \end{minipage}
    \hfill
    \begin{minipage}{0.48\textwidth}
        \centering
        \includegraphics[width=0.9\linewidth]{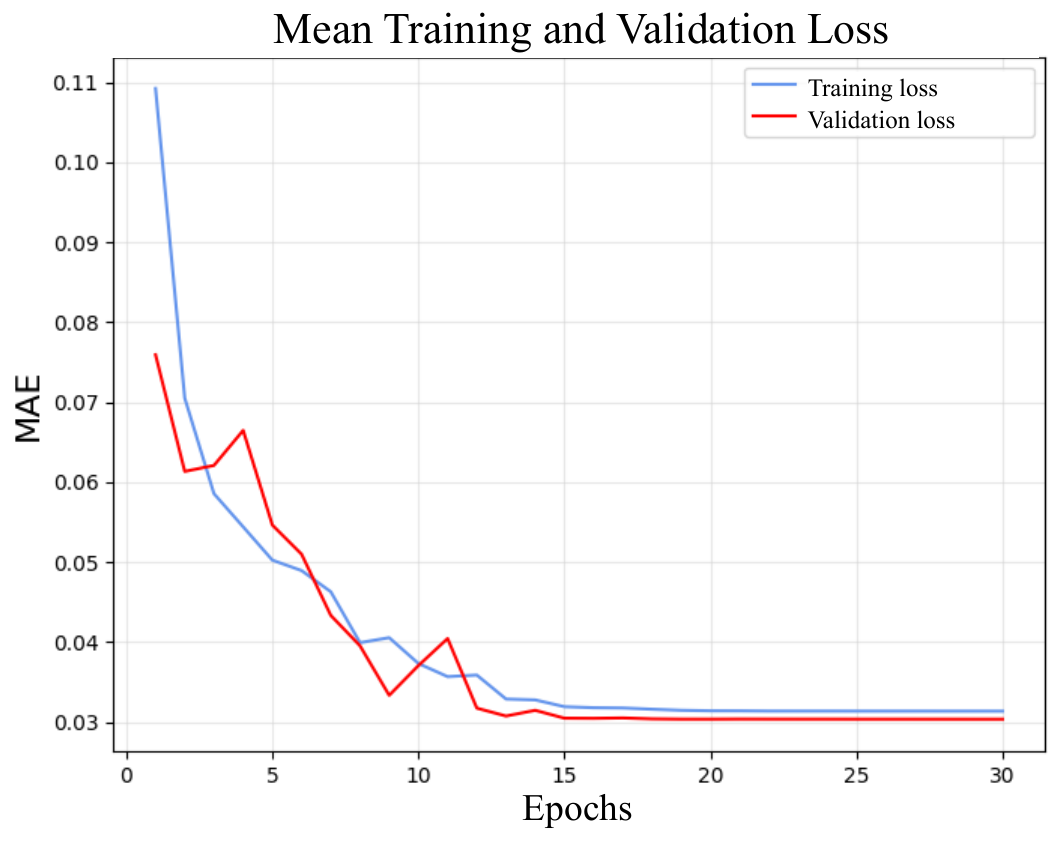}
        \caption*{(f) Experiment 2 – 5 layers.}
        \label{fig:q2_lossvento5}
    \end{minipage}

    \caption{ \justifying~Average values of the training loss (blue) and validation loss (red) over 30 training epochs, averaged across 10 runs per experiment. Subfigures (a), (c), and (e) show the results of Experiment 1 with 1, 3, and 5 variational layers, respectively. Subfigures (b), (d), and (f) correspond to Experiment 2.}
    \label{fig:q_lossvento}
\end{figure*}

Figure \ref{fig:q_lossvento} shows the average training and validation loss values, which indicate training efficiency and generalization ability, respectively. Finally, Figure \ref{fig:metricas_vento} compares the MAEs from each of the six QNN architectures evaluated in this section, along with the classical baseline.

\begin{table}[h!] 
\centering
\caption{ \justifying~  Experimental environment configuration for wind forecasting.}
\label{tab:config_vento}
\begin{tabular}{p{2cm}p{3.5cm}p{3cm}}
\toprule
\textbf{Element} & \textbf{Attribute} & \textbf{Description} \\
\midrule
Training set           & 361    & Number of data points used to train the model \\
Test set               & 5     & Number of data points used to test the model \\
Features               & 8      & Number of climate variables used as input to the model \\
Qubits                 & 8      & Number of qubits used to process the model \\
Epochs                 & 30     & Number of times the model iterates over the entire training set \\
Learning rate          & 0.1    & Learning rate used during training \\
Validation split       & 0.1    & Portion of training set used for validation \\
Batch size             & 10     & Number of samples processed before parameter updates \\
Hardware               & Nvidia Tesla V100 GPU (Kuatomu) & Hardware used to run the experiments \\
Software tools         & Pennylane, TensorFlow, Keras, Pandas, Scikit-learn, NumPy, Matplotlib & Classical and quantum tools used to build the model \\
\bottomrule
\end{tabular}
\end{table}


\subsubsection{Classical Model} \label{subsec2_c}

\begin{figure}[h!]
    \centering    \includegraphics[width=0.9\linewidth]{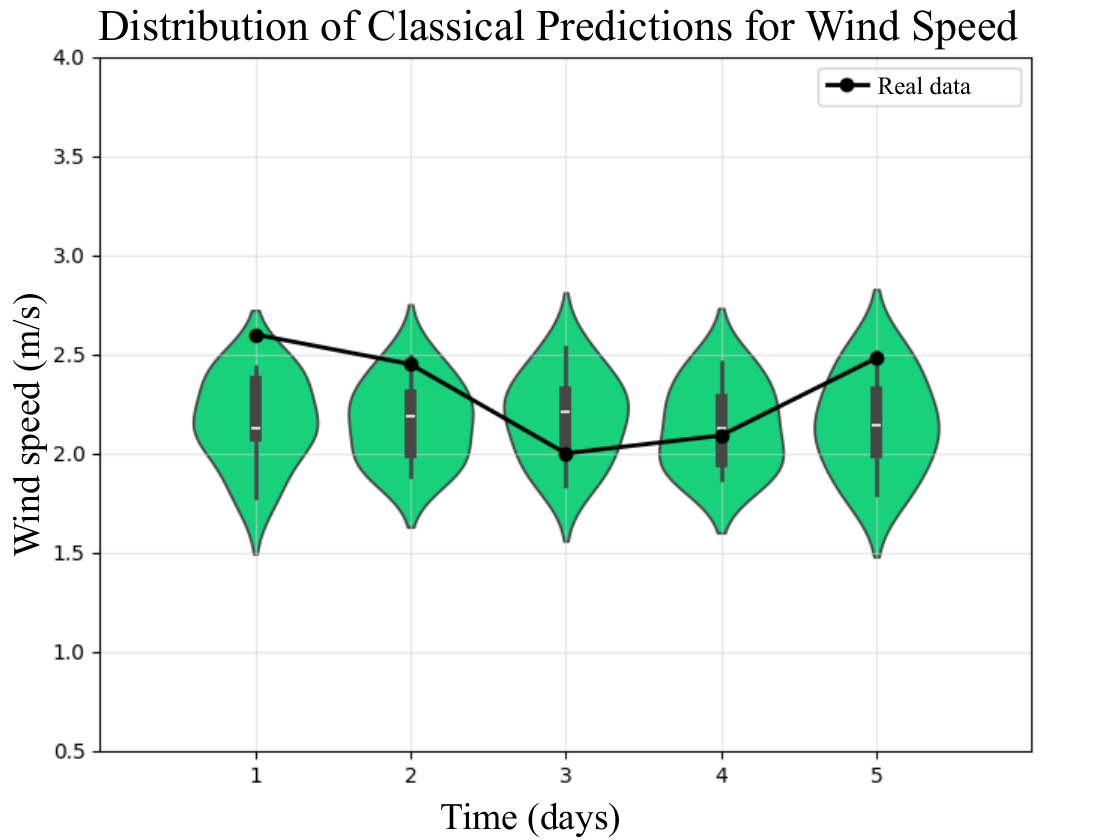}
    \caption{ \justifying~Distribution of classical wind speed forecasts for 5 days ahead, calculated from the average of 10 runs.}
    \label{fig:prevclassicavento_dist}
\end{figure}

\begin{figure}[h!]
    \centering
    \includegraphics[width=0.9\linewidth]{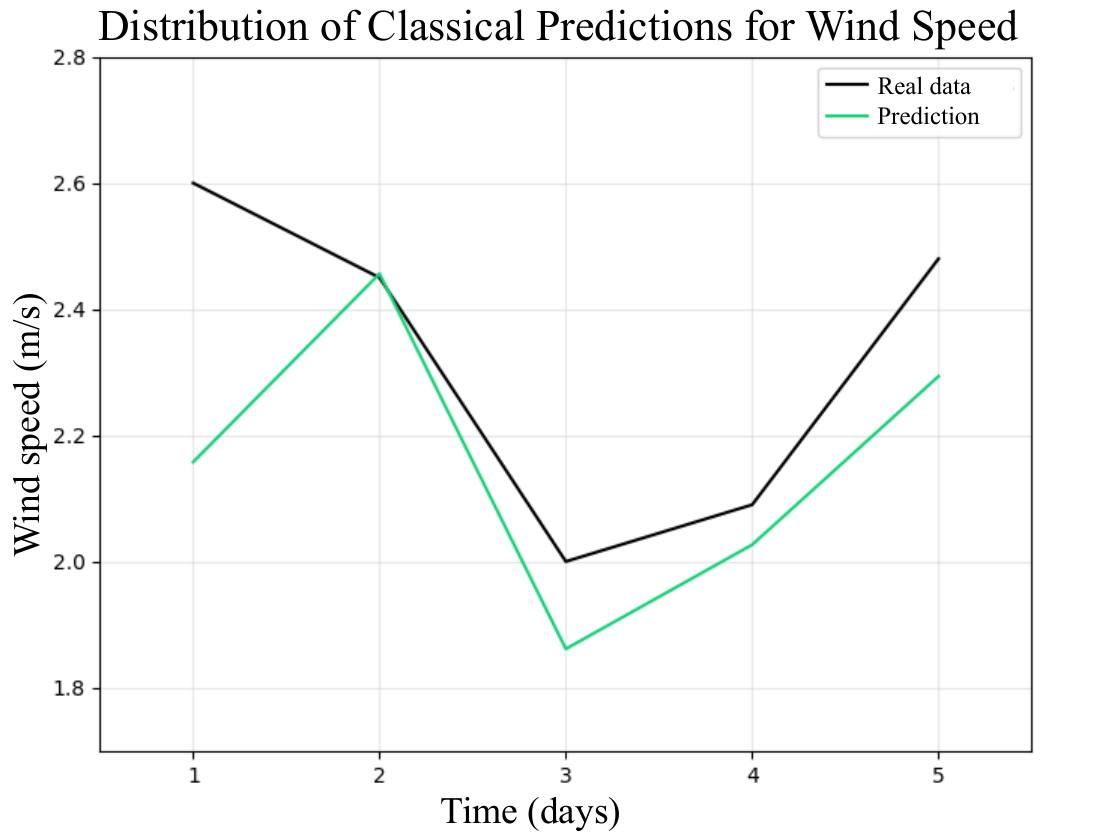}
    \caption{ \justifying~Mean classical wind speed forecast for 5 days ahead, calculated from the average of 10 runs.}
    \label{fig:prevclassicavento}
\end{figure}

\begin{figure}[h!]
    \centering
    \includegraphics[width=0.9\linewidth]{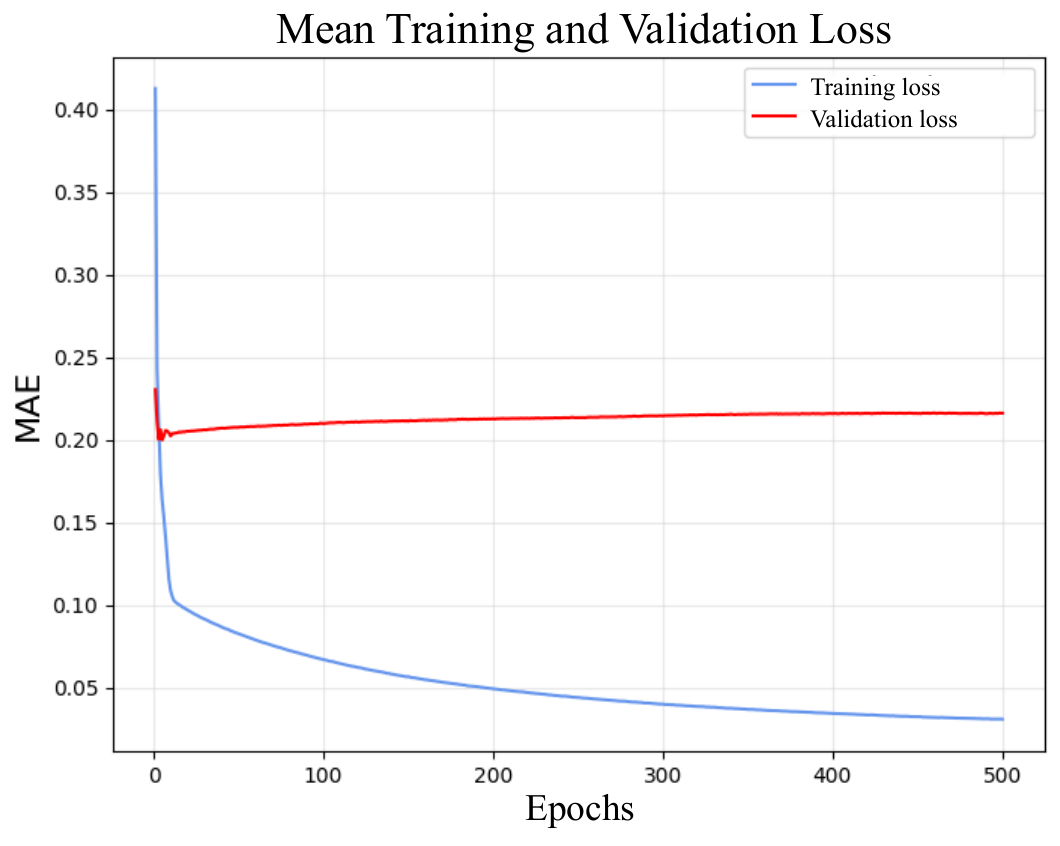}
    \caption{ \justifying~Average training loss (blue) and validation loss (red) for classical wind speed forecasting over 5 days, calculated from the average of 10 runs. A total of 500 epochs was used.}
    \label{fig:lossclassicavento}
\end{figure}

The RNN used for classical wind speed forecasting was configured with the same number of \textit{features}, $validation\_split$, $batch\_size$, \textit{hardware}, and training/testing proportions as the QNN, as shown in Table~\ref{tab:config_vento}. Similarly to the temperature prediction configuration, 256 neurons were used, and the initial learning rate was set to $lr = 0.001$. Additionally, the number of training epochs was also increased to 500 to analyze the behavior of the training and validation losses. Figure~\ref{fig:prevclassicavento_dist} shows the distribution of classical wind speed forecasts, followed by the corresponding mean values in Figure~\ref{fig:prevclassicavento}, and the average training and validation loss across 500 epochs in Figure~\ref{fig:lossclassicavento}.

As previously described, Figure \ref{fig:prevclassicavento_dist} displays violin plots encompassing the real data across all five forecast days. Notably, day 1 is the only case where the actual value lies outside the boxplot range, falling within the outlier region. The mean value obtained from the 10 algorithm runs is shown in Figure \ref{fig:prevclassicavento} and follows a pattern similar to the quantum forecast, in which the first day also exhibits the most significant discrepancy between predicted and actual values—already evident in the previous figure. The second forecast day yielded the most accurate result within the prediction window. From Day 2 to Day 5, the model demonstrated satisfactory performance, closely following the downward and upward trends in the data. Finally, Figure \ref{fig:lossclassicavento} revealed training loss convergence at 0.03 and validation loss convergence at 0.22. While the training loss exhibited a consistent decay, the validation loss showed a slight upward trend, indicating that the model was well-fitted to the seen data and, as expected, had a comparatively lower but stable performance on the unseen validation set.

\subsubsection*{Comparison between predictive models}

Unlike the temperature forecasting task, the classical model (RNN) generally yielded lower variance in wind speed prediction, indicating greater robustness. On the other hand, the quantum model exhibited boxplots more closely aligned with the actual data, capturing patterns that the classical model appeared to overlook.

\begin{figure}[h!]
    \centering
    \includegraphics[scale=0.4]{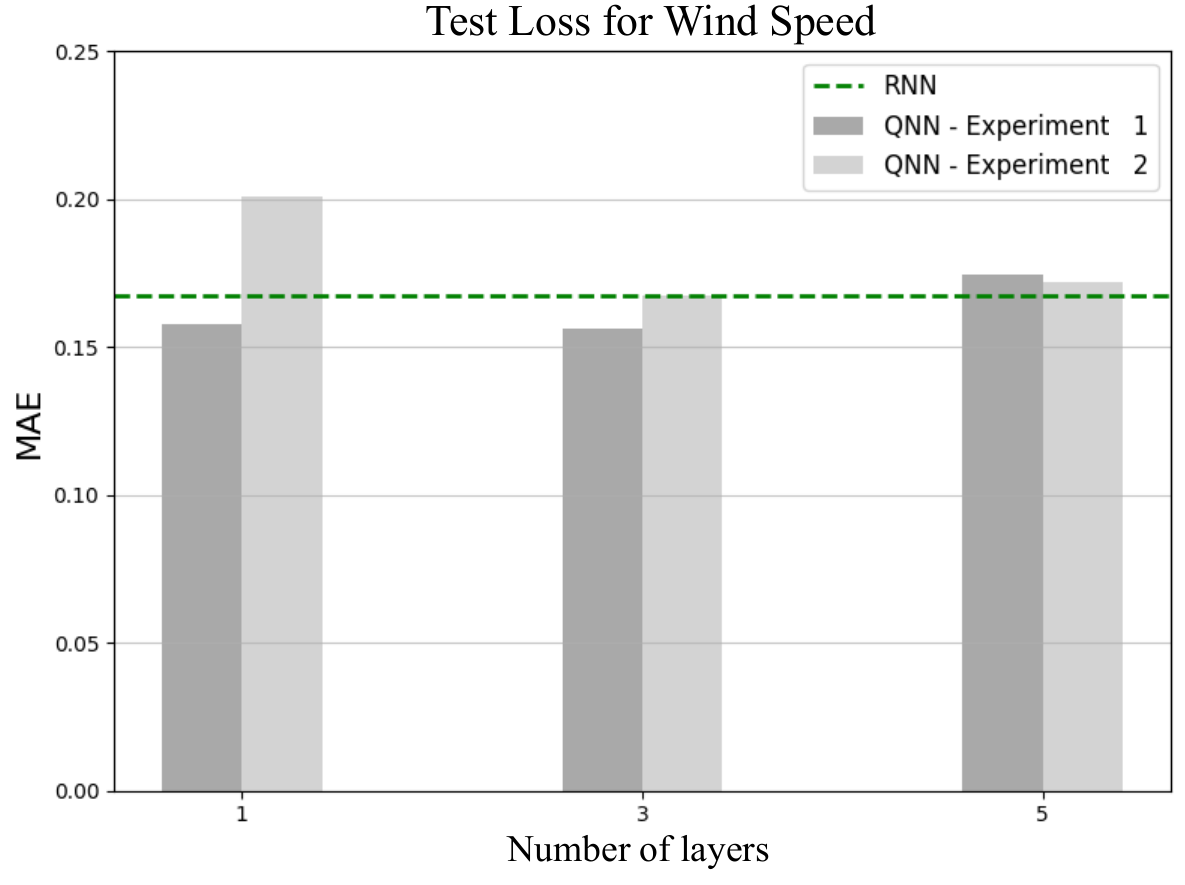}
    \caption{ \justifying~Mean absolute error (MAE) associated with the loss functions on the test set for wind speed forecasting. The dashed green line indicates the MAE achieved by the RNN; the dark gray bars represent the MAEs from Experiment 1 of the QNN, and the light gray bars correspond to the MAEs from Experiment 2 of the QNN. The analysis was performed for 1, 3, and 5 layers.}
    \label{fig:metricas_vento}
    \vspace{0.5cm}
\end{figure}

The MAE and accuracy metrics achieved by the predictive models for wind speed forecasting are summarized in Table \ref{tab:metricas_vento} and illustrated in Figure \ref{fig:metricas_vento}.

\begin{table}[h!] 
\centering
\caption{ \justifying~  Performance metrics (MAE and Accuracy) obtained by the predictive models QNN and RNN for wind speed forecasting.}
\label{tab:metricas_vento}
\begin{tabular}{@{}llccc@{}}
\toprule
\textbf{Model} & \textbf{Experiment} & \textbf{Layer} & \textbf{MAE} & \textbf{Accuracy (\%)} \\
\midrule
\multirow{3}{*}{QNN} 
 & \multirow{3}{*}{Experiment 1} & 1 & 0.158 & 84.2 \\
 &                                & 3 & 0.156 & 84.4 \\
 &                                & 5 & 0.174 & 82.6 \\
\cmidrule(lr){2-5}
 & \multirow{3}{*}{Experiment 2} & 1 & 0.201 & 79.9 \\
 &                                & 3 & 0.168 & 83.2 \\
 &                                & 5 & 0.172 & 82.8 \\
\midrule
\multirow{1}{*}{RNN} 
 & - & - & 0.167 & 83.3 \\
\bottomrule
\end{tabular}
\end{table} 







As in the previous section, the MAE achieved by the RNN exceeded half of the QNN simulations, this time with a value of 0.167. Although the quantum model did not exhibit linear behavior, both experiments achieved better performance with a network structure comprising three variational layers. QNN Experiment 1 depicted the lowest MAE with three variational layers. This result is reinforced by the analysis of Figure~\ref{fig:q_lossvento} (c), where the validation loss exhibited the lowest initial MAE among all QNN architectures, later showing the smoothest decay.


\section{Conclusion}\label{sec4}

This study explored the application of quantum machine learning (QML) to short-term and medium-term climate forecasting—an emerging frontier in artificial intelligence. A comprehensive theoretical foundation was established, and a practical implementation was conducted using daily meteorological data from Barreiras over one year. Two predictive models, a quantum neural network (QNN) and a classical recurrent neural network (RNN), were compared using standardized datasets from NASA POWER.

Simulation results revealed that the classical RNN consistently achieved intermediate performance relative to QNN simulations. QNN architectures produced the most accurate forecasts: a single-layer \textit{StronglyEntanglingLayer} QNN achieved 69.6\% accuracy for temperature prediction, while a three-layer \textit{EntanglingLayer} QNN reached 84.4\% accuracy for wind speed forecasting. However, the nonlinear behavior observed in QNN performance across varying depths suggests that additional factors beyond entanglement play significant roles in training efficiency and accuracy.

Notably, the QNN demonstrated reduced dispersion in temperature prediction, whereas the RNN was more stable for wind speed data. These findings suggest that classical models may perform better on datasets with lower variability, while quantum models are potentially more robust to noisy, high-variance data. QNNs also showed better alignment with real data distributions and greater adaptability to abrupt trend shifts—traits desirable in extreme event forecasting. Furthermore, QNN training with temperature data converged more rapidly than with wind data, aligning with expectations given the greater instability of wind patterns.

Lastly, the main contribution of this work lies in demonstrating the practical feasibility and potential advantages of QNNs in climate forecasting. By incorporating Pearson correlation analysis for feature selection and performing comparative evaluations across six QNN architectures and a classical RNN, this study contributes to the gradual evolution of QML implementations, fostering greater accuracy and applicability as the technology matures.

Future work should investigate statistical preprocessing techniques to mitigate the influence of high-variance values and enhance model stability. Additionally, understanding why wind speed prediction yielded higher accuracy than temperature prediction, despite its inherent instability, warrants further investigation—potentially through analysis using data from other geographic regions. The limitations of the NASA POWER dataset, including its daily resolution and non-geostationary satellite sources, may also impact forecasting precision and warrant attention in subsequent studies.\\

\textbf{Funding.}
This work has been fully funded by the project ``iNOVATeQ Lato Sensu Specialization in Quantum Computing – Researcher'' supported by Quantum Industrial Innovation (QuIIN) - EMBRAPII CIMATEC Competence Center in Quantum Technologies, with financial resources from the PPI IoT/Manufatura 4.0 of the MCTI grant number 053/2023, signed with EMBRAPII. C. Cruz, G.F. de Jesus, and M.H.F. da Silva thank the Bahia State Research Support Foundation (FAPESB) for financial support (grant numbers APP0041/2023 and PPP0006/2024). \\

\textbf{Supplementary information.}
The source codes and implementation details used in this study are available at \cite{maheloisa_weather_2025} to ensure reproducibility and transparency.\\

\textbf{Acknowledgements.}
The authors thank the Latin American Quantum Computing Center (LAQCC) - SENAI CIMATEC for access to the KUATOMU quantum simulator, which was used in obtaining results.\\

\textbf{Conflicts of interest.}
The authors have no conflicts to disclose.\\


\bibliography{apssamp}

\end{document}